\documentstyle[12pt]{article}
\include{epsf}
\begin{document}

\title{\Large \bf Classical technical analysis of  Latin American market indices.
Correlations in Latin American currencies ($ARS$, $CLP$, $MXP$) exchange rates
with respect to $DEM$, $GBP$, $JPY$ and $USD$\footnote{Happy Birthday, Dietrich;
by now you should be rich !}}

\author{ \large \bf M. Ausloos$^1$ and K. Ivanova$^2$ \\ $^1$ GRASP and SUPRAS,
B5, Sart Tilman, \\ B-4000 Li\`ege, Euroland\\$^2$ Pennsylvania State University,
\\ University Park, PA 16802, USA \\ }



\maketitle

\begin{abstract}

The classical technical analysis methods of financial time series based on the
moving average and momentum is recalled. Illustrations use the IBM share price
and Latin American (Argentinian MerVal, Brazilian Bovespa and Mexican IPC) market
indices. We have also searched for scaling ranges and exponents in exchange rates
between Latin American currencies ($ARS$, $CLP$, $MXP$) and other major
currencies $DEM$, $GBP$, $JPY$, $USD$, and $SDR$s. We have sorted out
correlations and anticorrelations of such exchange rates with respect to $DEM$,
$GBP$, $JPY$ and $USD$. They indicate a very complex or speculative behavior.

\end{abstract}

\vskip 2cm


\newpage

\section{Introduction}

The buoyancy of the US dollar is a reproach to stagnant Japan, recessing Europe
economy and troubled developing countries like Brazil or Argentina. Econophysics
aims at introducing statistical physics techniques and physics models in order to
improve the understanding  of financial and economic matters. Thus when this
understanding is established, econophysics might $later$ help in the well being
of humanity. In so doing several techniques have been developed to analyze the
correlations of the fluctuations of stocks or currency exchange rates. It is of
interest to examine cases pertaining to rich or developing economies.

In the first sections of this report we  recall the classical technical analysis
methods of stock evolution. We recall the notion of moving averages and
(classical) momentum.  The case of IBM and Latin American market indices serve as
illustrations.

In 1969 the International Monetary Fund created the {\it special drawing rights}
$SDR$, an artificial currency defined as a basket of national currencies $DEM$,
$FRF$, $USD$, $GBP$ and $JPY$. The $SDR$ is used as an international reserve asset,
to supplement members existing reserve assets (official holdings of gold, foreign
exchange, and reserve positions in the IMF). The $SDR$ is the IMF's unit of
account. Four countries maintain a currency peg against the $SDR$. Some private
financial instruments are also denominated in $SDR$s.\cite{SDR1,SDR2} Because of
the close connections between the developing countries and the IMF, we search for
correlations between the fluctuations of $ARS$, $CLP$ and $MXP$ exchange rates
with respect to $SDR$ and the currencies that form this artificial money. In the
latest sections of this report, we compare the correlations of such fluctuations
as we did in our previous results on $EUR$ exchange rates fluctuations with
respect to $USD$, $GBP$ and $JPY$ \cite{maki3,kimalg,tokyokima}.

\section{Technical Analysis: IBM and Latin America Markets}

Technical indicators as {\it moving average} and $momentum$ are part of the
classical technical analysis and much used in efforts to predict market movements
\cite{Achelis}. One question is whether these techniques provide adequate ways to
read the trends.

Consider a time series $x(t)$ given at N discrete times $t$. The series (or
signal) moving average $M_{\tau}(t)$ over a time interval $\tau$ is defined as

\begin{equation} M_{\tau}(t)=\frac{1}{\tau}\sum_{i=t}^{t+\tau-1} x(i-\tau) \qquad
t=\tau+1,\dots,N \end{equation} i.e. the average of $x$ over the last $\tau$ data
points. One can easily show that if the signal $x(t)$ increases (decreases) with
time, $M_{\tau}(t)<x(t)$ ($M_{\tau}(t)>x(t)$). Thus, the moving average captures
the trend of the signal given the period of time $\tau$. The IBM daily closing
value price signal between Jan 01, 1990 and Dec 31, 2000 is shown in Fig. 1 (top
figure) together with Yahoo moving average taken for $\tau=50$~days
\cite{yahooibm}. The bottom figure shows the daily $volume$ in millions.

\begin{figure}[ht] \begin{center} \leavevmode \epsfysize=9cm
\epsffile{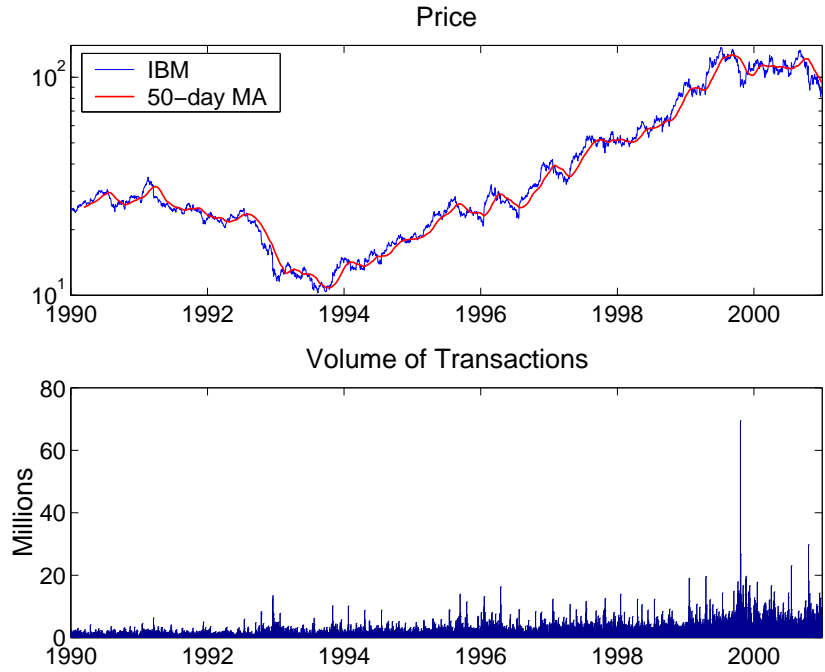}
\caption{IBM daily
closing value signal between Jan 01, 1990 and Dec 31, 2000, i.e. 2780 data points
with Yahoo moving average for $\Delta T$ = 50 day. (top); the bottom figure shows
the daily volume} \label{eps1}\end{center} \end{figure}

There can be as many trends as moving averages as $\tau$ intervals. The shorter
the $\tau$ interval the more sensitive the moving average. However, a too short
moving average may give false messages about the long time trend of the signal.
In Fig. 2(a) two moving averages of the IBM signal for $\tau$=5 days (i.e. 1
week) and 21~days (i.e. 1 month) are compared for illustration.

\begin{figure} 
\begin{center} \leavevmode \epsfysize=5.1cm\epsffile{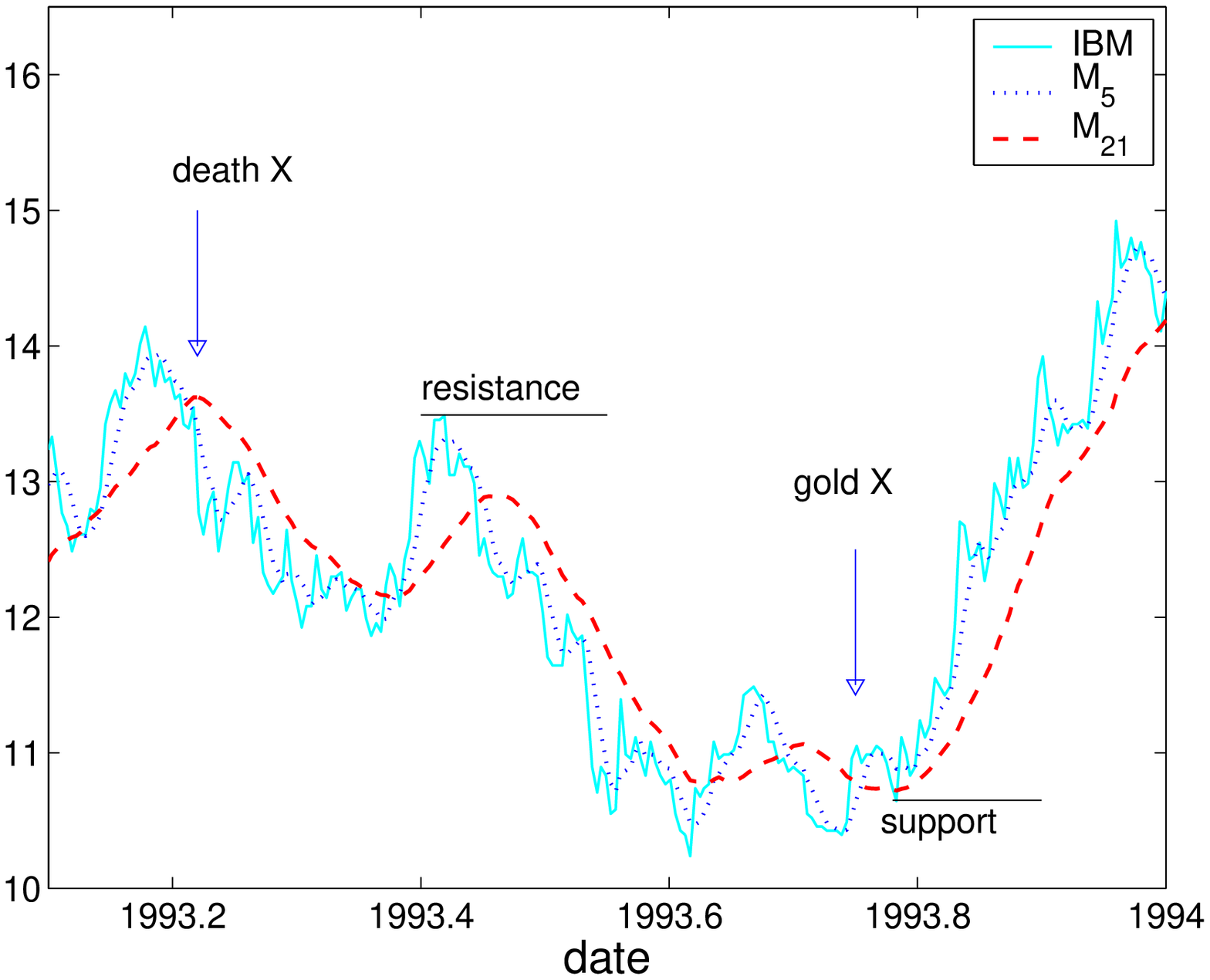}
\hfill \leavevmode \epsfysize=5.4cm\epsffile{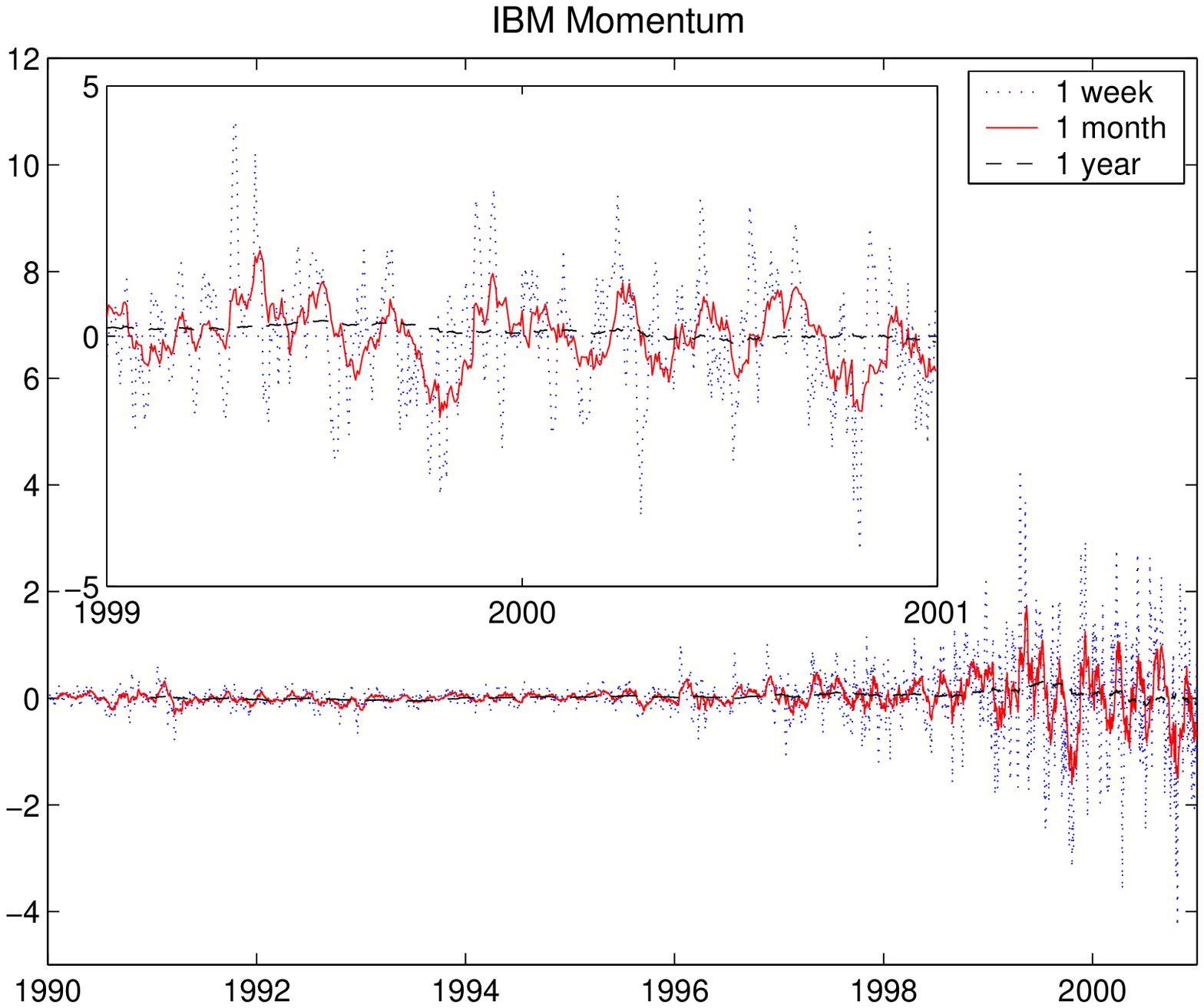} \vfill
\leavevmode \epsfysize=5.2cm\epsffile{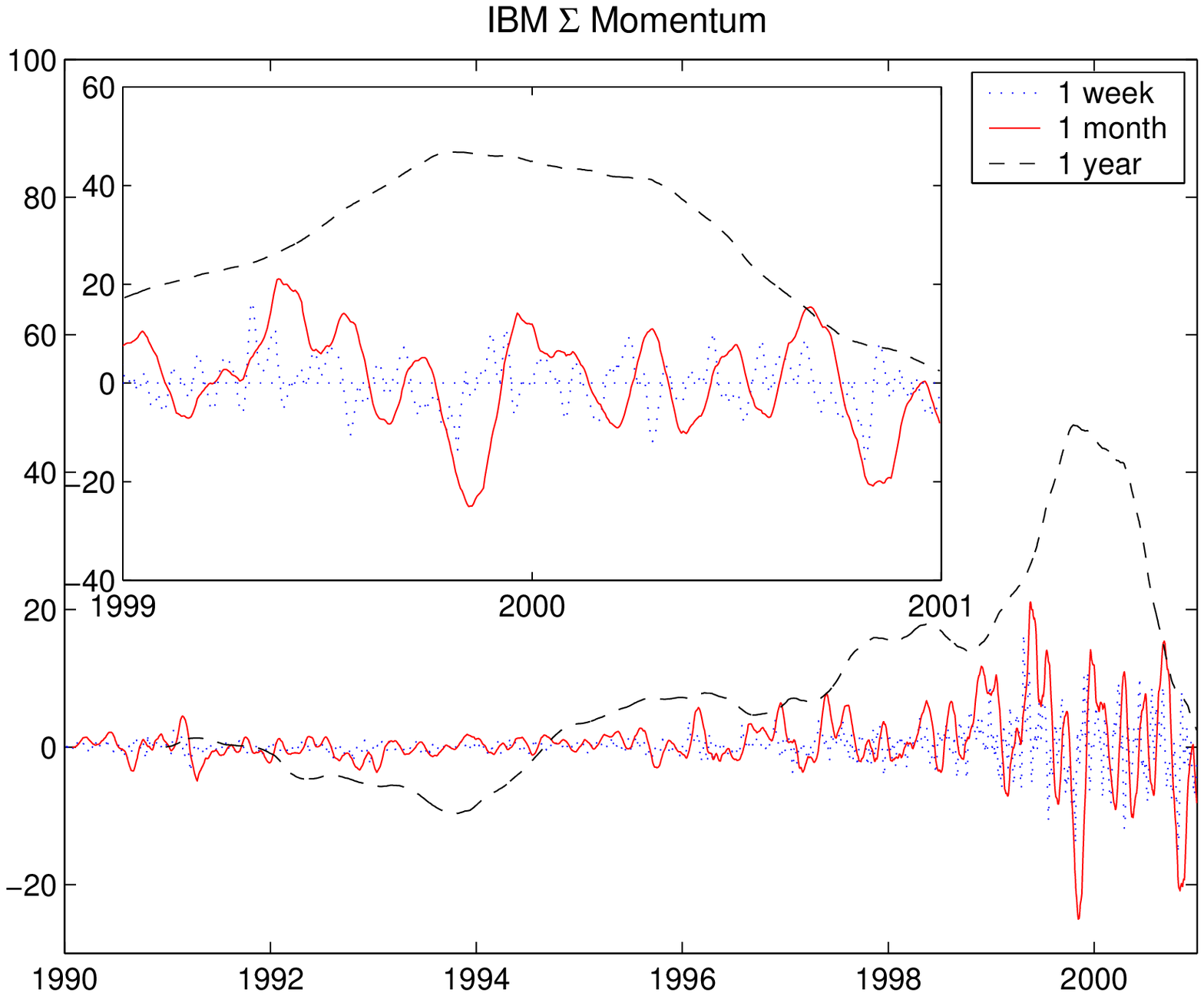} \hfill
\leavevmode \epsfysize=5.2cm\epsffile{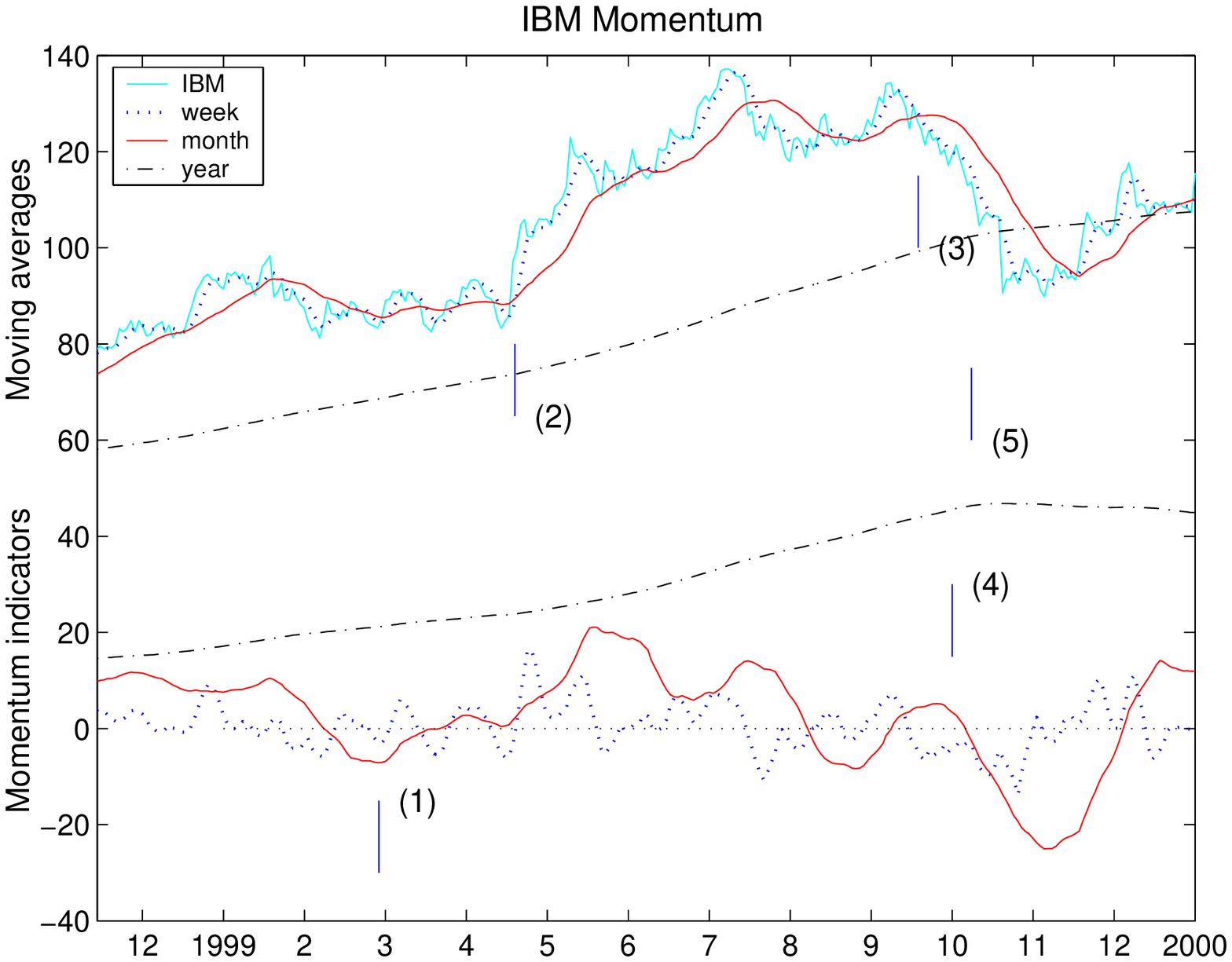} \vfill \caption[]{(a) IBM daily
closing value signal between Jan 01, 1990 and Dec 31, 2000, with two moving
averages, $M_{\tau_1}$ and $M_{\tau_2}$ for $\tau_1=5$~days and $\tau_2=21$~days.
(b) IBM classical momentum (c) moving average of IBM classical momentum for three
different time periods $\tau$, 1 week (dot curve), 1 month (solid curve) and 1
year (dot-dash curve). (d) IBM signal and the six market indicators for three
time horizons, short-term (weekly) (dot curve), medium-term (monthly) (solid
curve) and log-term (yearly) (dot-dash curve). IBM original signal and its moving
averages are divided by 10 for readability} \label{eps2}\end{center} \end{figure}

The intersections of the price signal with a moving average can define so-called
lines of resistance or support \cite{Achelis}. A line of resistance is observed
when the local maximum of the price signal $x(t)$ crosses a moving average
$M_{\tau}(t)$. A support line is defined if the local minimum of $x(t)$ crosses
$M_{\tau}(t)$. In Fig. 2(a) lines of resistance happen around May 1993 and lines
of support around Sept 1993. Support levels indicate the price where the majority
of investors believe that prices will move higher, and resistance levels indicate
the price at which a majority of investors feel prices will move lower. Other
features of the trends are the intersections between $two$ moving averages
$M_{\tau_1}$ and $M_{\tau_2}$ which are usually due to drastic changes in the
trend of $x(t)$ \cite{nvmama}. Consider two moving averages of IBM price signal
for $\tau_1=5$~days and $\tau_2=21$~days (Fig. 2(a)). If $x(t)$ increases for a
long period of time before decreasing rapidly, $M_{\tau_1}$ will cross
$M_{\tau_2}$ from above. This event is called a ''death cross'' in empirical
finance \cite{Achelis}. In contrast, when $M_{\tau_1}$ crosses $M_{\tau_2}$ from
below, the crossing point coincide with an upsurge of the signal $x(t)$. This
event is called a ''gold cross''. Therefore, it is of interest to study the
density of crossing points between two moving averages as a function of the size
difference of the $\tau$'s defining the moving averages. Based on this idea, a
new and efficient approach has been suggested in Ref.\cite{nvmama} in order to
estimate an exponent that characterizes the roughness of a signal.

The so called $momentum$ is another instrument of the technical analysis and we
will refer to it here as the {\it classical momentum}, in contrast to the
generalized momentum \cite{GenMom}. The classical momentum of a stock is defined
over a time interval $\tau$ as

\begin{equation} R_{\tau}(t)=\frac{x(t)-x(t-\tau)}{\tau}= \frac{\Delta x}{\Delta
t} \qquad t=\tau+1,\dots,N \end{equation}

The momentum $R_{\tau}$ for three time intervals, $\tau=5, 21$ and 250~days, i.e.
one week, one month and one year, are shown in Fig. 2(b) for IBM. The longer the
period the smoother the momentum signal. Much information on the price trend
turns is usually considered to be found in a {\it moving average of the momentum}

\begin{equation} R^{\Sigma}_{\tau}(t) = \sum_{i=t}^{t+\tau-1}
\frac{x(i)-x(i-\tau)}{\tau} \qquad t=\tau+1,\dots,N \end{equation}

Moving averages of the classical momentum over 1 week, 1 month and 1 year for the
IBM price difference {\it over the same time intervals}, $R^{\Sigma}_{\tau}$ are
shown in Fig. 2(c). In Fig. 2(d) the IBM signal and its weekly (short-term),
monthly (medium-term) and yearly (long-term) moving averages are compared to the
weekly (short-term), monthly (medium-term) and yearly (long-term) momentum
indicators in order to better observe the bullish and bearish trends in 1999.

The message that is coming out of reading the combination of the these six
indicators states that one could start buying at the momentum bottom, as it is
for both monthly and weekly momentum indicators around mid February 1999 and buy
the rest of the position when the price confirms the momentum uptrend and rises
above the monthly moving average which is around March 1999. The first selling
signal is given during the second half of July 1999 by the death cross between
short and medium term moving averages and by the maximum of the monthly momentum,
which indicates the start of a selling. At the beginning of October 1999, occurs
the maximum of the long-term momentum. It is recommended that one can sell the
rest of the position since the price is falling down below the moving average.
Hence, it is said that {\it momentum indicators lead the price trend}. They give
signals before the price trend turns over.

Along the lines of the above for IBM, we analyze three Latin America financial
indices, Argentinian MerVal, Brazilian Bovespa and Mexican IPC (Indice de Precios
y Cotizaciones) applying the moving averages and the classical momentum concepts.
In Fig. 3 the time evolution of the MerVal stock over the time interval mid 1996
- mid 2001, is plotted with a simple moving average for $\tau=$50~days showing
the medium range trend of the price. The moving average   and classical momentum
of the MerVal stock prices for time horizons equal to one week, one month and one
year are shown Fig. 3.

\begin{figure} \begin{center} \leavevmode \epsfysize=9cm\epsffile{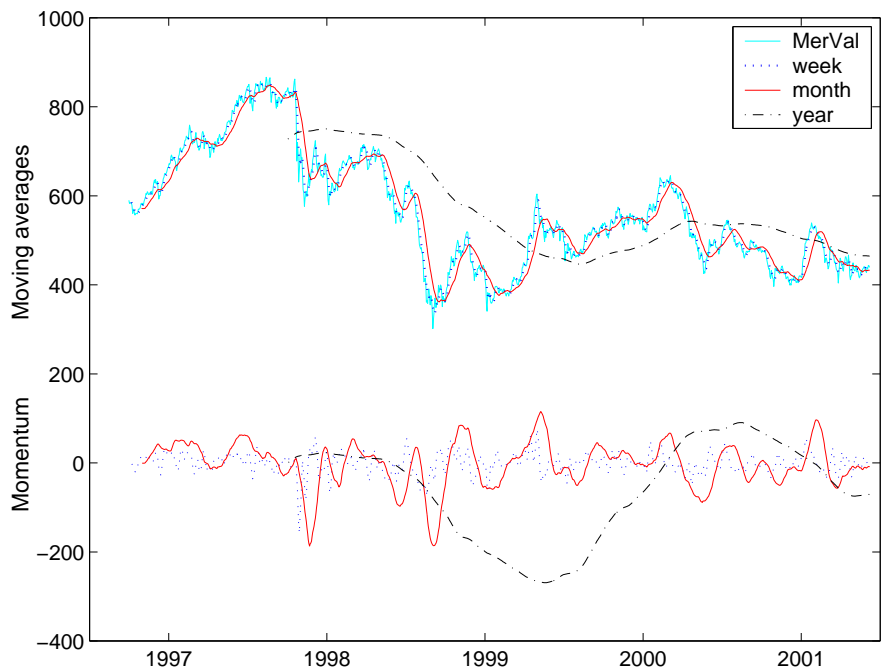}
\caption[]{ Argentinian MerVal daily closing value signal between Oct 08, 1996
and Jun 06, 2001, i.e. 1162 data points; classical momentum for three time
horizons, short-term (weekly) (dot curve), medium-term (monthly) (solid curve)
and log-term (yearly) (dot-dash curve)} \label{eps3}\end{center} \end{figure}

The cases of the Brazilian Bovespa and Mexican IPC (Indice de Precios y
Cotizaciones) are shown in Figs. 4  and 5  respectively.

\begin{figure} \begin{center} \leavevmode \epsfysize=9cm\epsffile{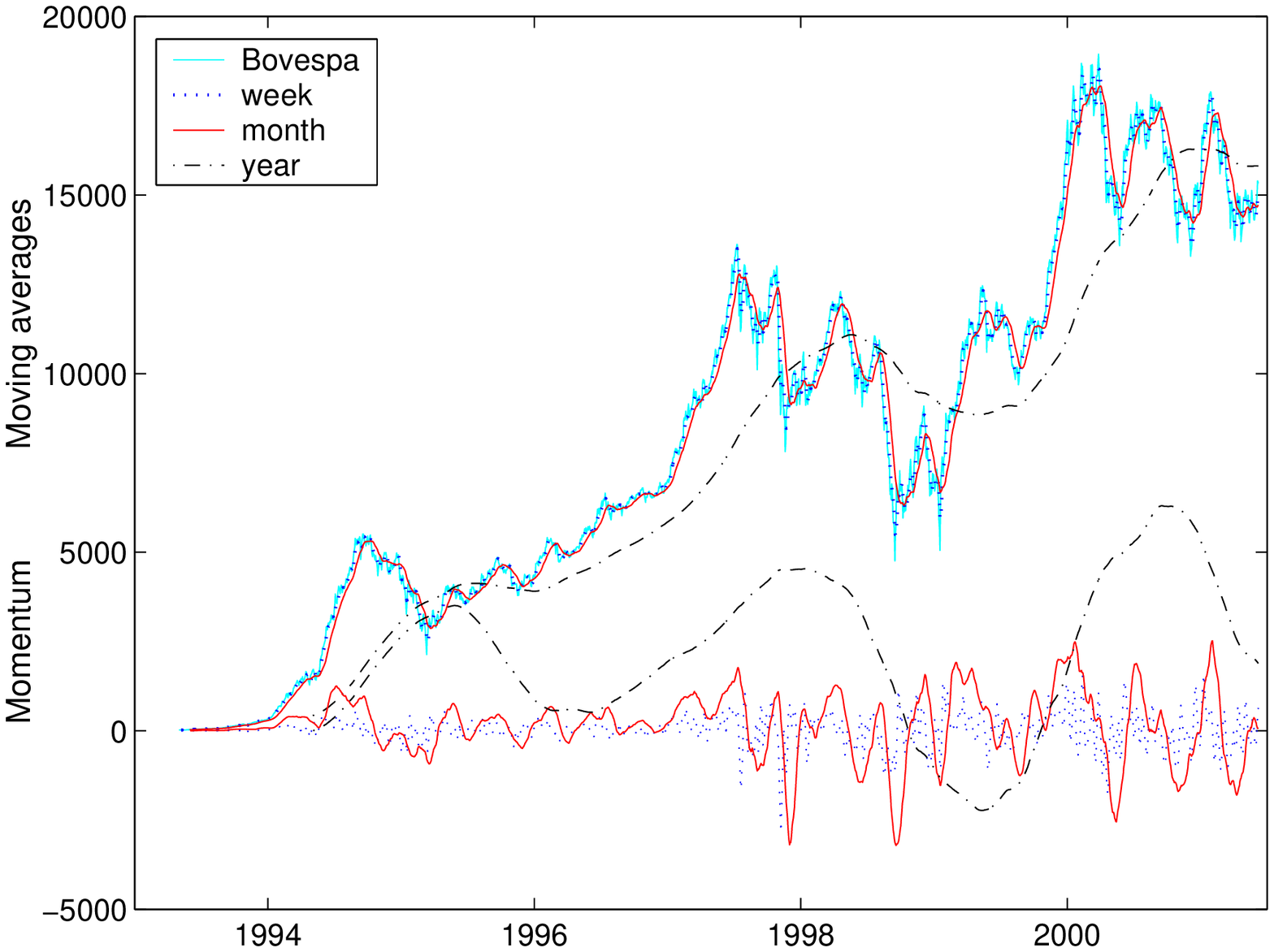}
\caption[]{ Brazilian Bovespa daily closing value signal between April 27, 1993
and Jun 06, 2001, i.e. 2010 data points; classical momentum for three time
horizons, short-term (weekly) (dot curve), medium-term (monthly) (solid curve)
and log-term (yearly) (dot-dash curve)} \label{eps4} \end{center}\end{figure}

\begin{figure} \begin{center} \leavevmode \epsfysize=9cm\epsffile{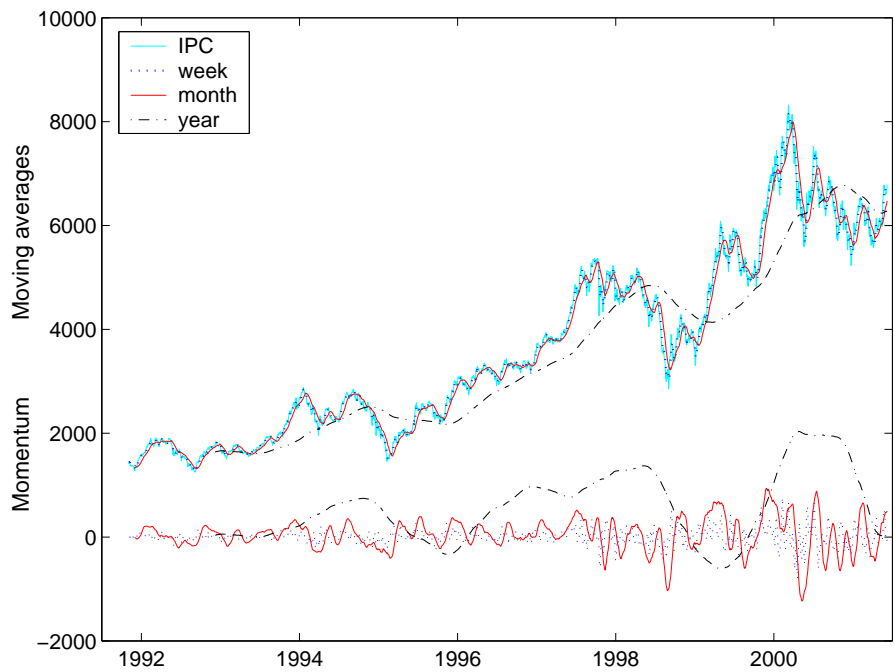}
\caption[]{ Mexican IPC daily closing value signal between Nov. 08, 1991 and Jun
06, 2001, i.e. 2384 data points; classical momentum for three time horizons,
short-term (weekly) (dot curve), medium-term (monthly) (solid curve) and log-term
(yearly) (dot-dash curve)} \label{eps5}\end{center} \end{figure}

\section{Exchange Rates and Special Drawing Rights}

In 1969 the International Monetary Fund (IMF) created the Special Drawing Rights
(ticker symbol $SDR$; sometimes $XDR$ is also used), an artificial currency unit
defined a basket of national currencies. The $SDR$ is used as an international
reserve asset, to supplement members' existing reserve assets (official holdings
of gold, foreign exchange, and reserve positions in the IMF). The $SDR$ is the
IMF's unit of account: IMF voting shares and loans are all denominated in $SDR$s.
The $SDR$ serves as the unit of account for a number of other international
organizations, including the WB. Four countries maintain a currency peg against
the $SDR$. Some private financial instruments are also denominated in $SDR$s.

The basket is reviewed every five years to ensure that the currencies included in
the basket are representative of those used in international transactions and
that the weights assigned to the currencies reflect their relative importance in
the world's trading and financial systems. Following the completion of the most
recent regular review of $SDR$ valuation on October 11, 2000, the IMF's Executive
Board agreed on changes in the method of valuation of the $SDR$ and the
determination of the $SDR$ interest rate, effective Jan. 01, 2001.

The $SDR$ artificial currency can be represented as an weighted sum of the five
currencies $C_i$, $i= 1, 5$:

\begin{equation} 1 SDR = \sum_{i=1}^{5} \gamma_i \, C_i \end{equation}

\noindent where $\gamma_i$ are the currencies weights in percentage (Table 1) and
$C_i$ denote the respective currencies, U.S. Dollar ($USD$), German Mark ($DEM$),
French Franc ($FRF$), Japanese Yen ($JPY$), British Pound ($GBP$).
\begin{table} \begin{center} \caption{Currency Weights in $SDR$ Basket (In
Percent)} \begin{tabular}{ccc} \hline Currency &Last Revision&Revision of\\
&January 1, 2001 &January 1, 1996\\ $USD$ &45&39\\ $EUR$ &29&\\ $DEM$&&21\\
$FRF$&&11\\ $JPY$&15&18\\ $GBP$&11&11\\ \hline \end{tabular} \end{center}
\label{table1}\end{table}

On January 1, 1999, the German Mark and French Franc in the $SDR$ basket were
replaced by equivalent amounts of $EUR$. The relevant exchange rates (ExR) are
shown in Figs. 6-8.
\begin{figure} \begin{center} \leavevmode \epsfysize=3.3cm\epsffile{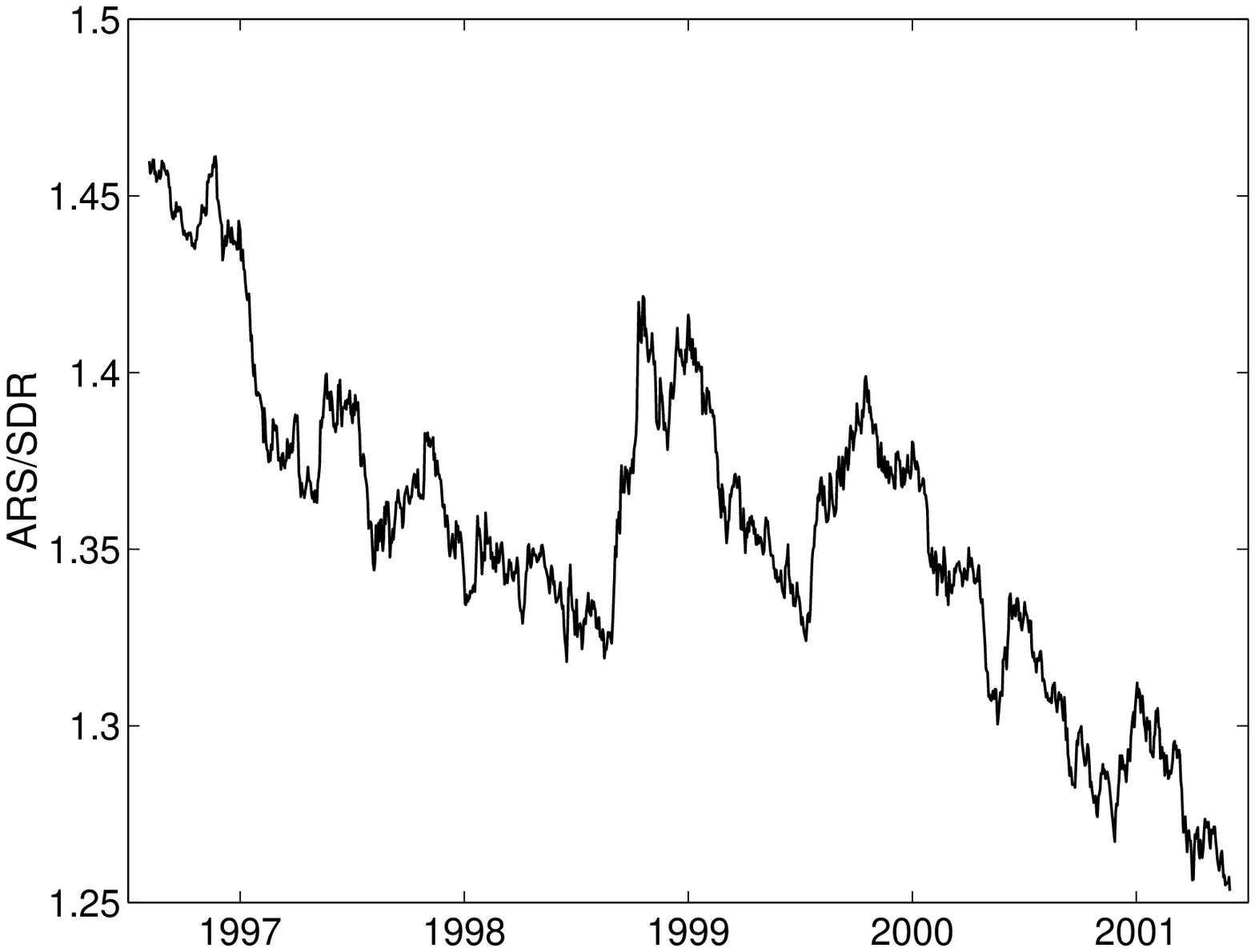} \hfill
\leavevmode \epsfysize=3.3cm\epsffile{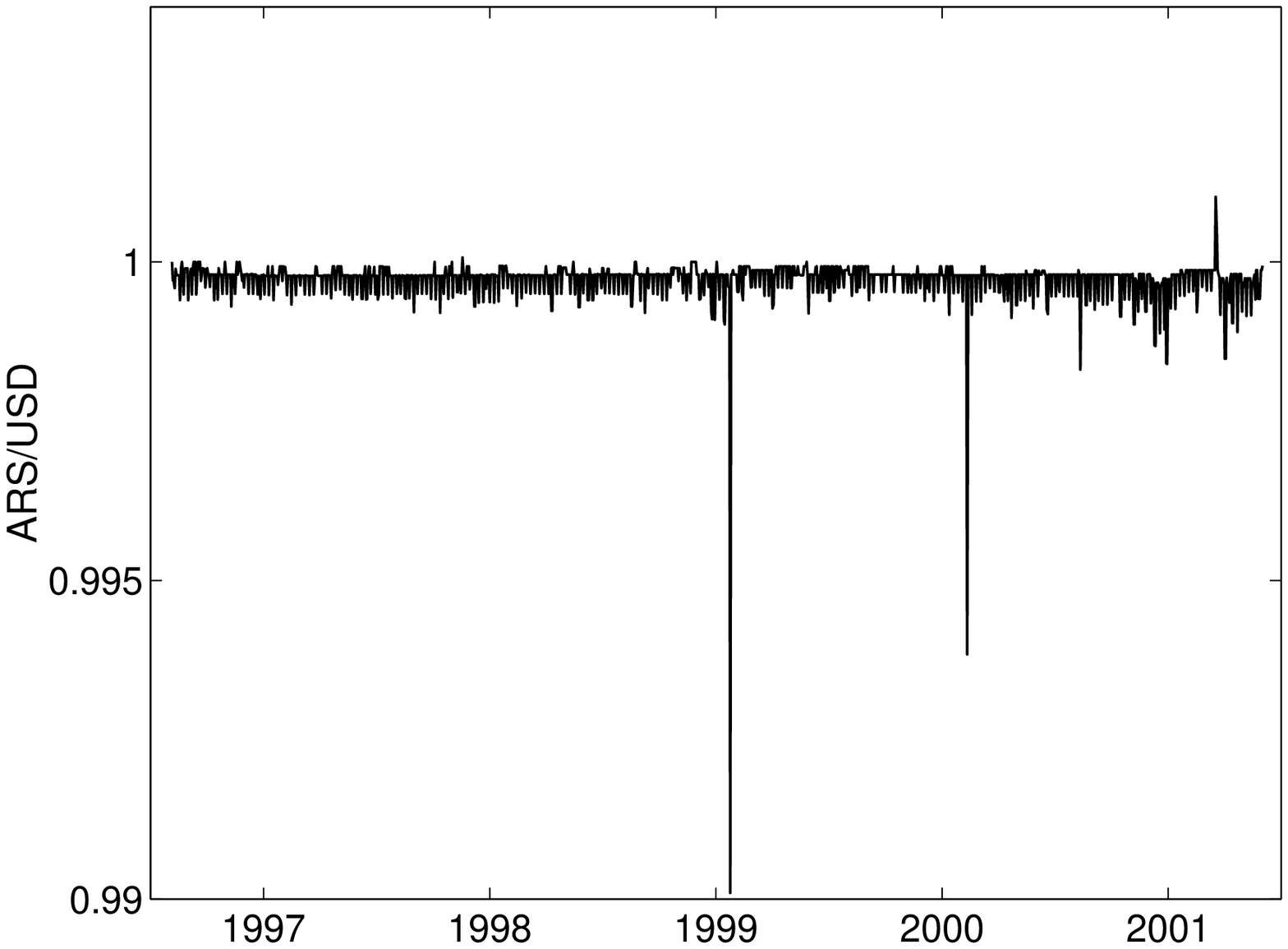} \hfill
\leavevmode \epsfysize=3.3cm\epsffile{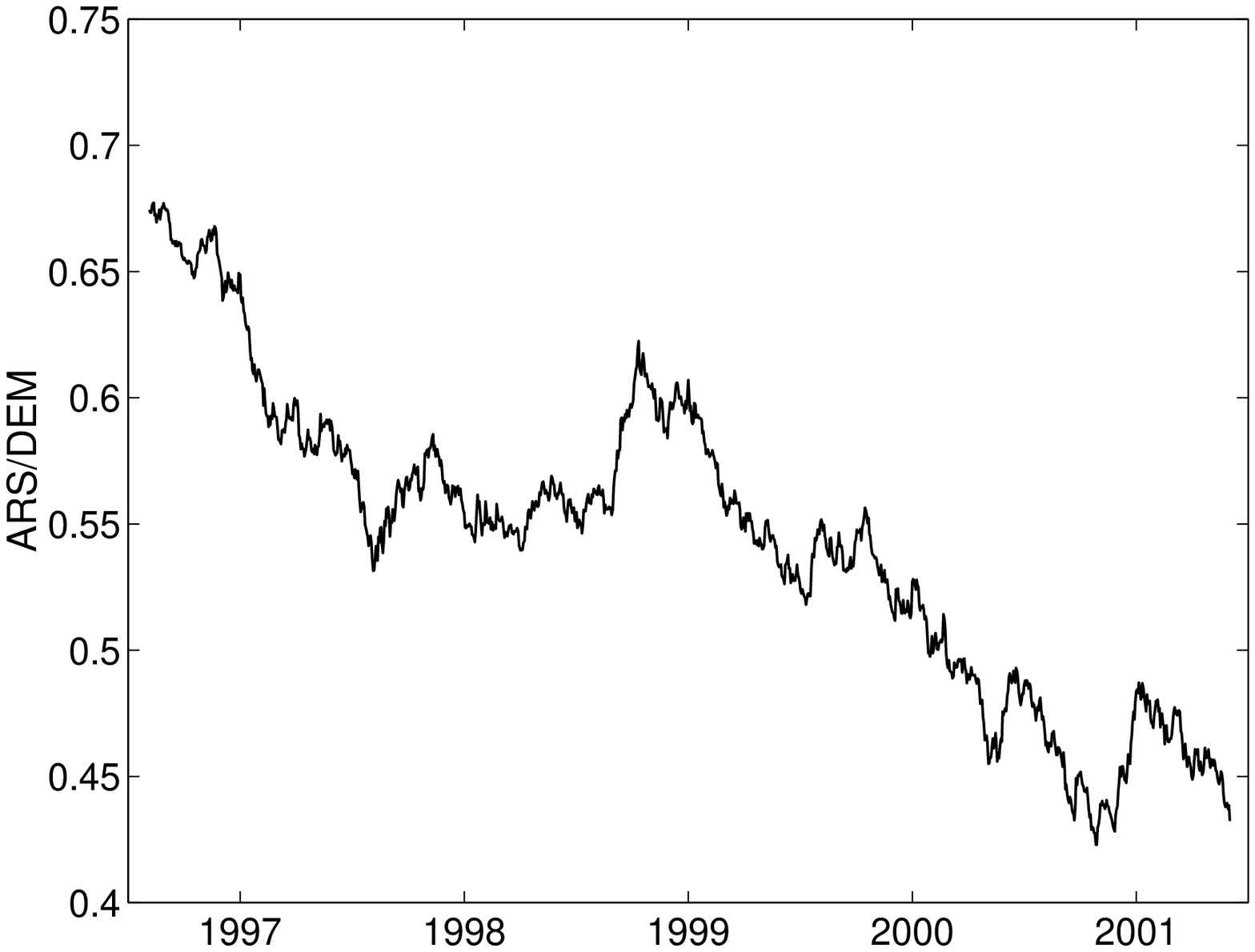} \vfill
\leavevmode \epsfysize=3.3cm\epsffile{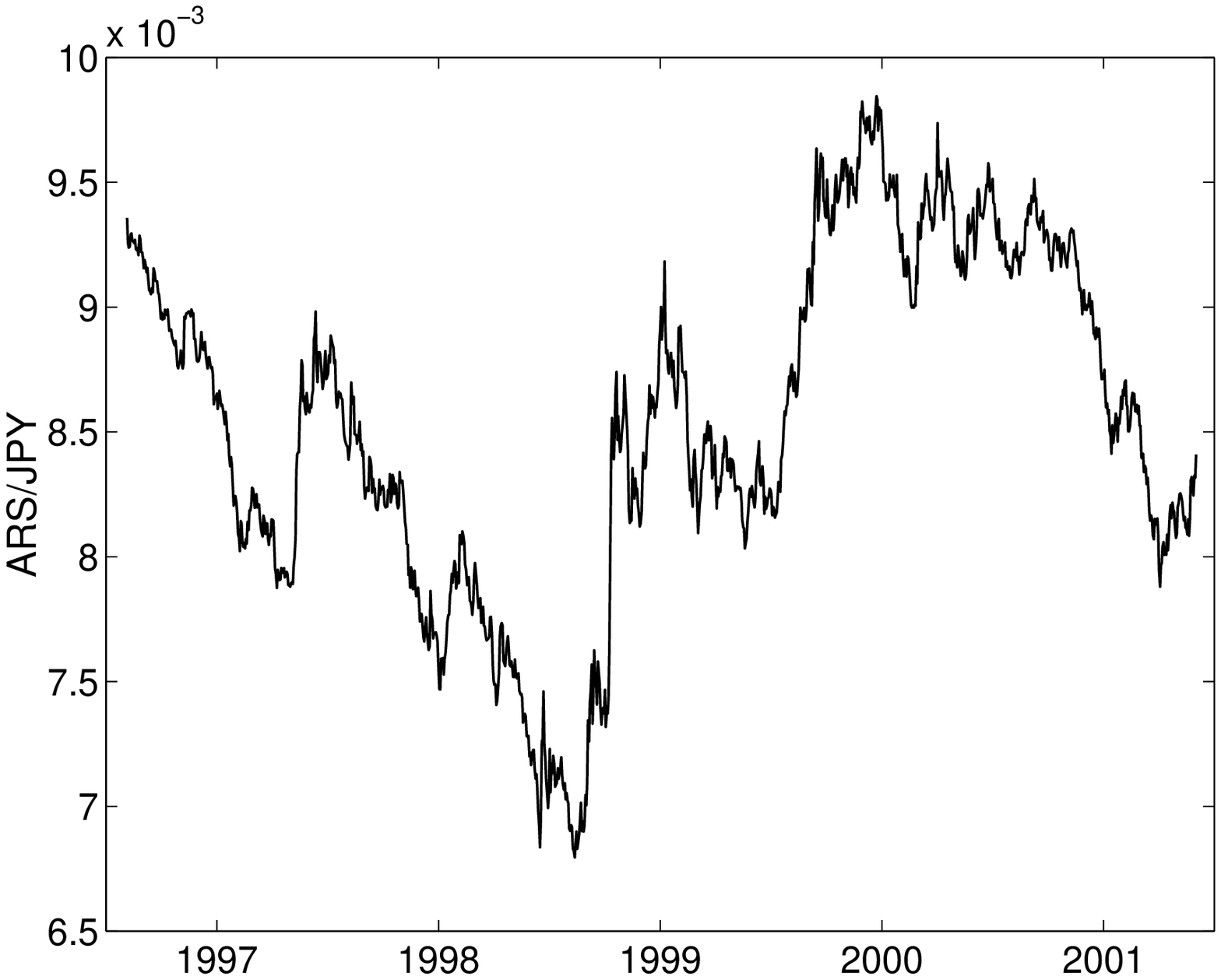} \hfill
\leavevmode \epsfysize=3.3cm\epsffile{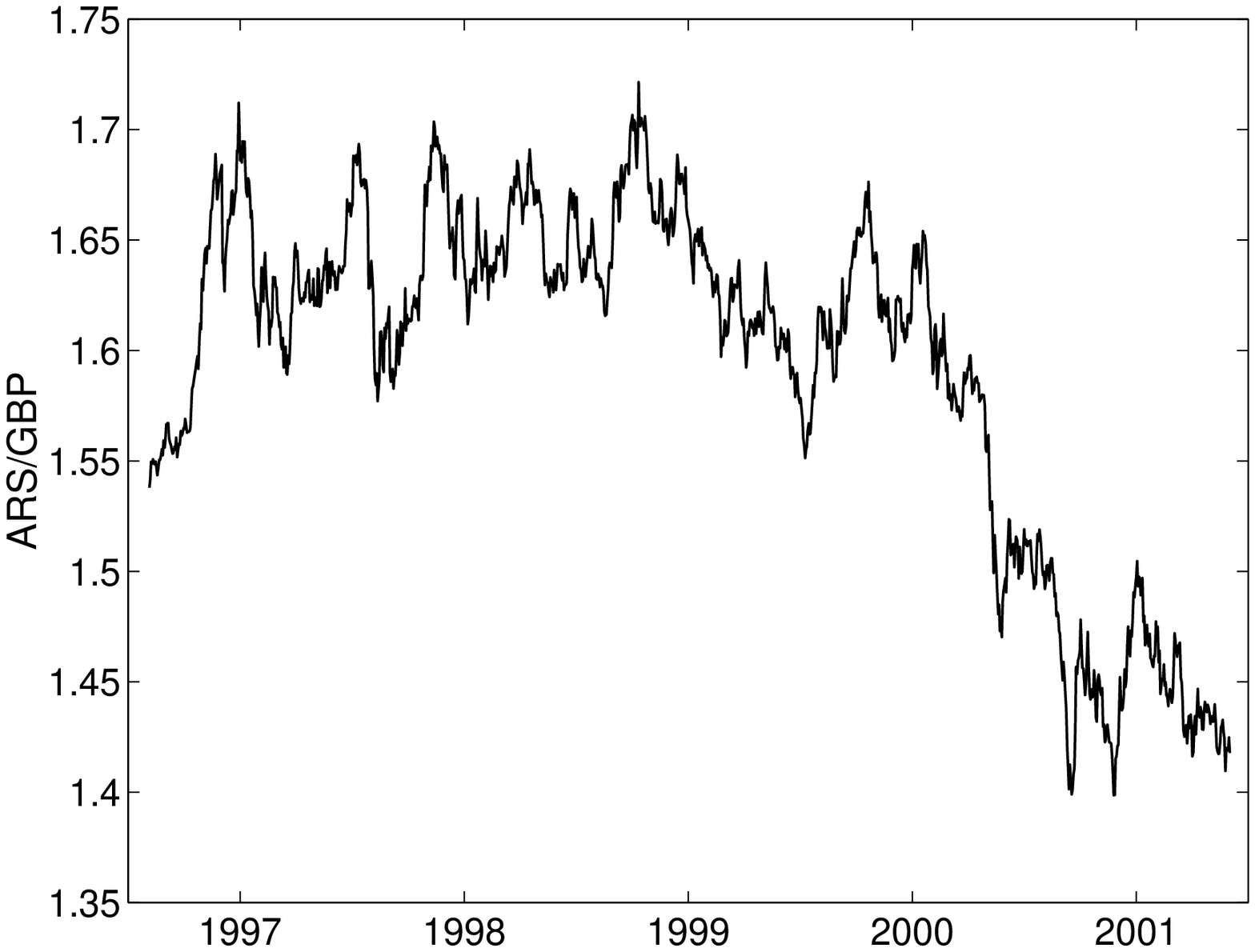} \vfill \caption[]{Exchange rates
of $ARS$ with respect to $SDR$, $USD$, $DEM$, $JPY$, $GBP$ for the time interval
between Aug. 6, 1996 and May 31, 2001, i.e. 1208 data points, as available on
http://pacific.commerce.ubc.ca/xr/ website } \label{eps6}\end{center} \end{figure}
\begin{figure}\begin{center}\leavevmode \epsfysize=3.3cm\epsffile{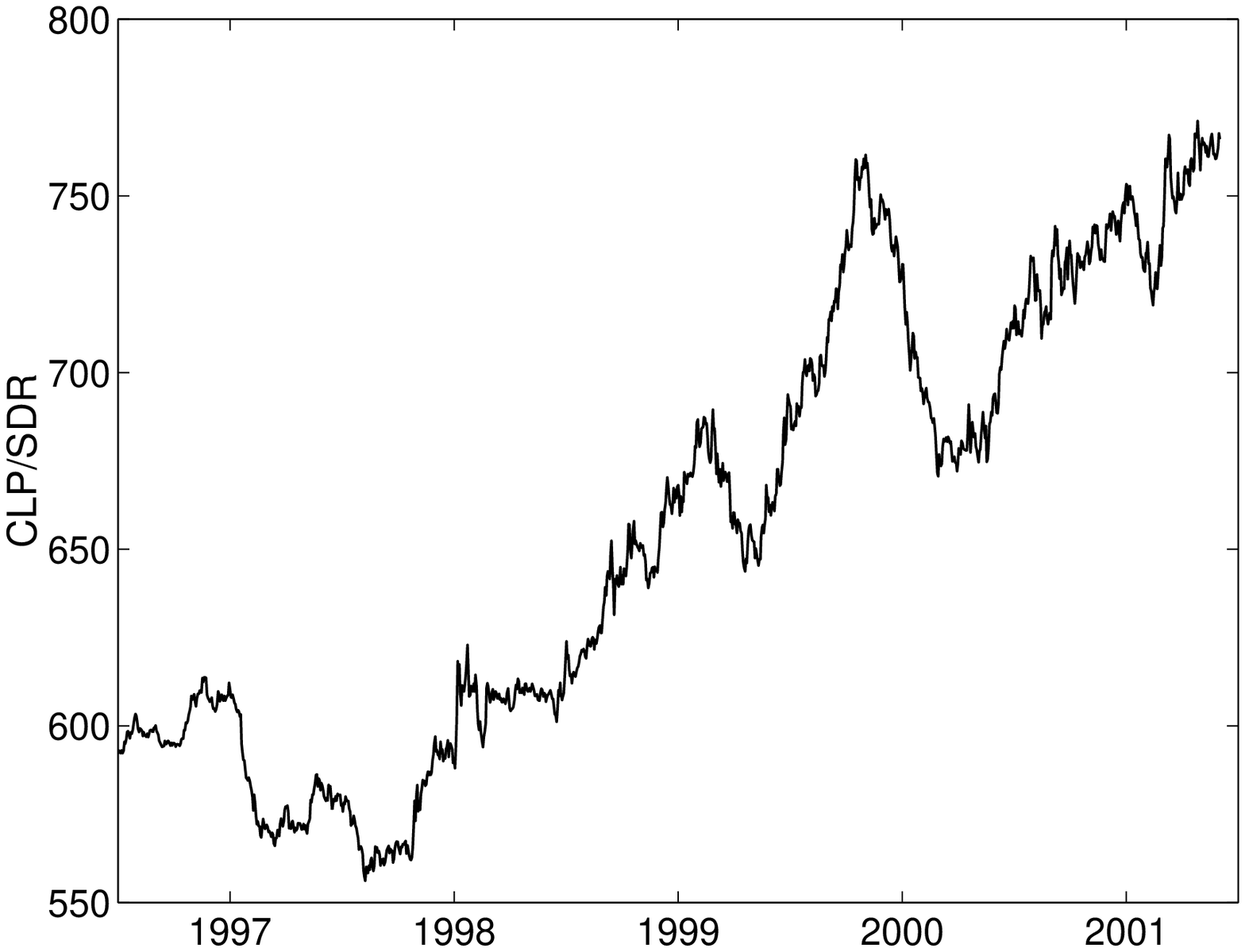} \hfill
\leavevmode \epsfysize=3.3cm\epsffile{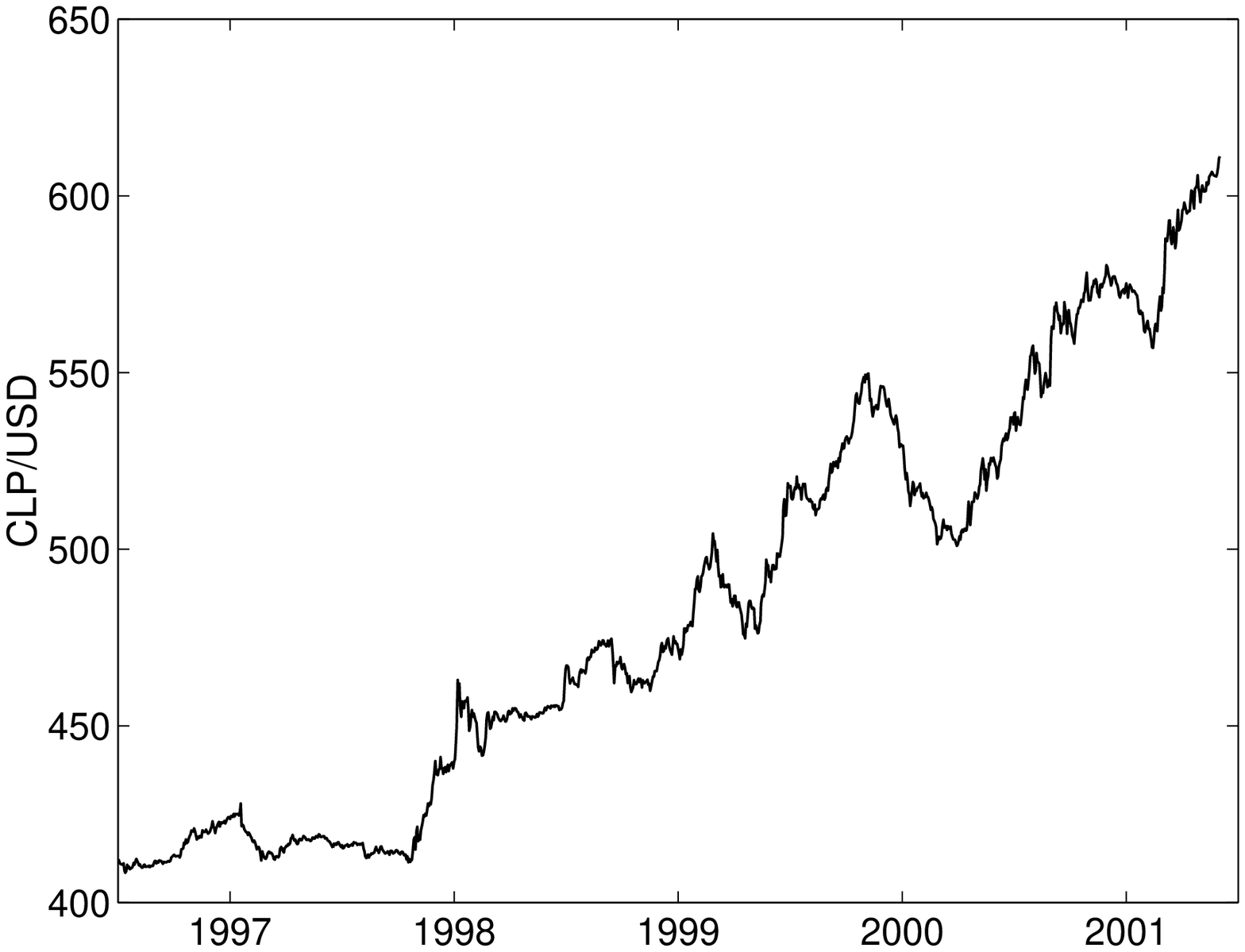} \hfill
\leavevmode \epsfysize=3.3cm\epsffile{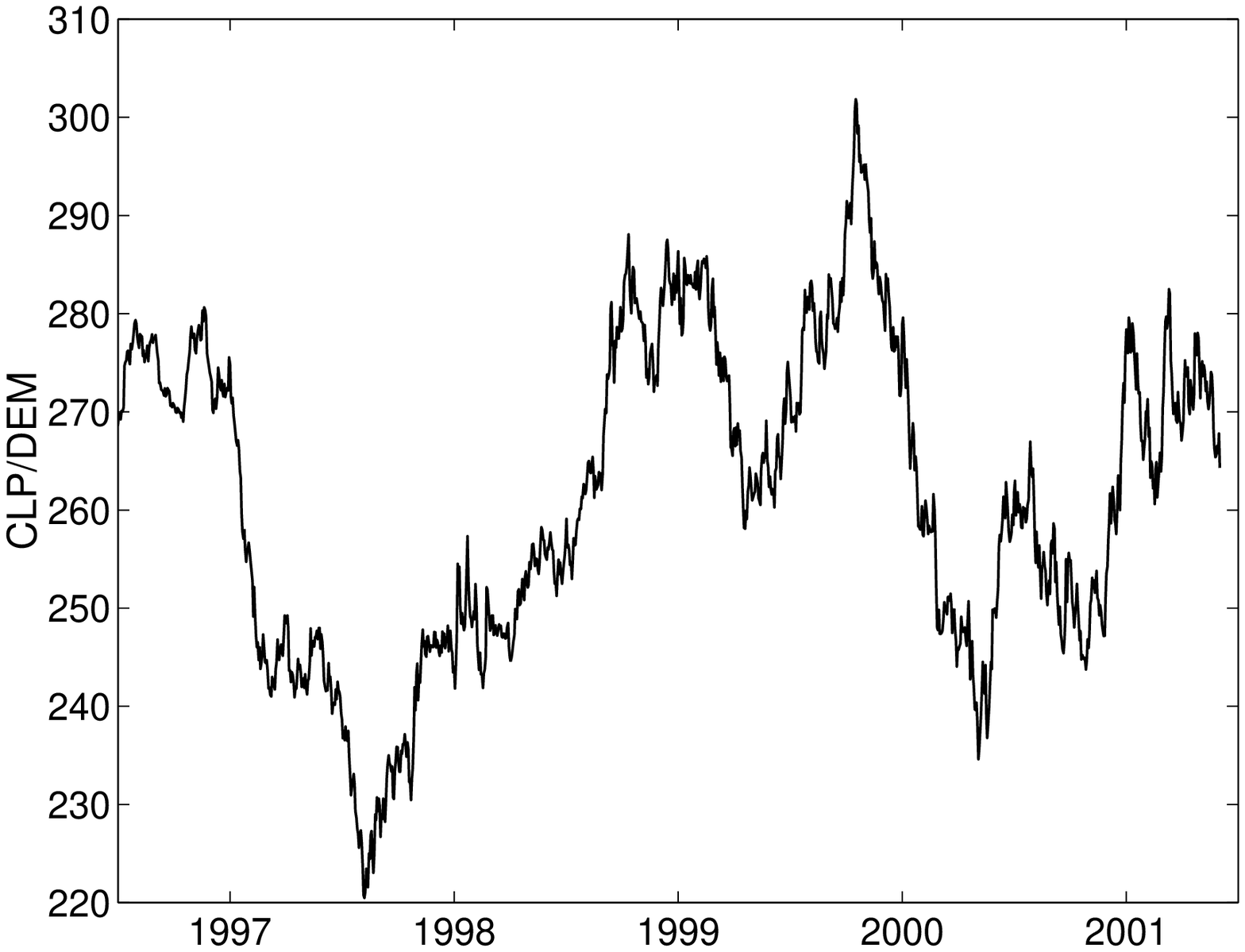} \vfill
\leavevmode \epsfysize=3.3cm\epsffile{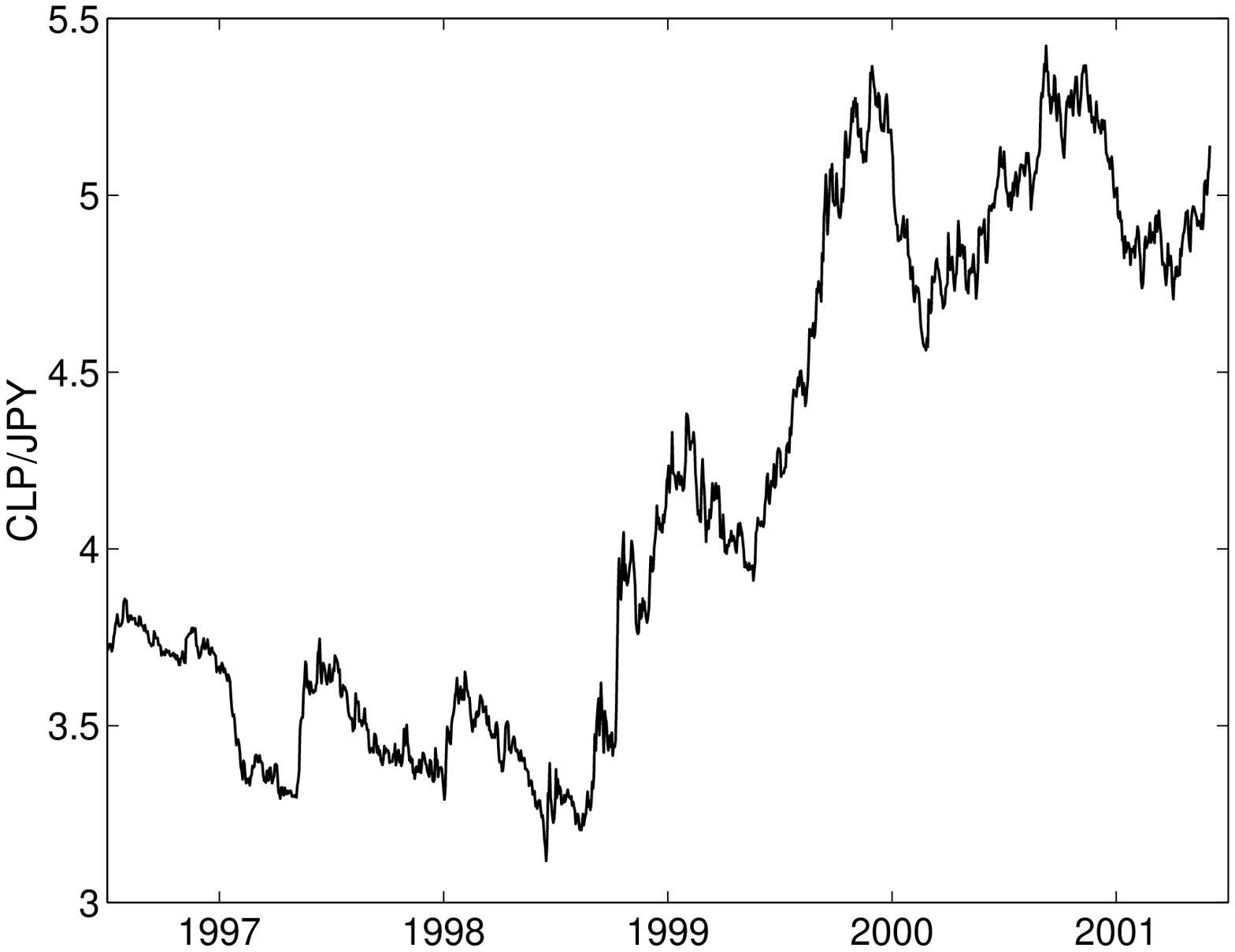} \hfill
\leavevmode \epsfysize=3.3cm\epsffile{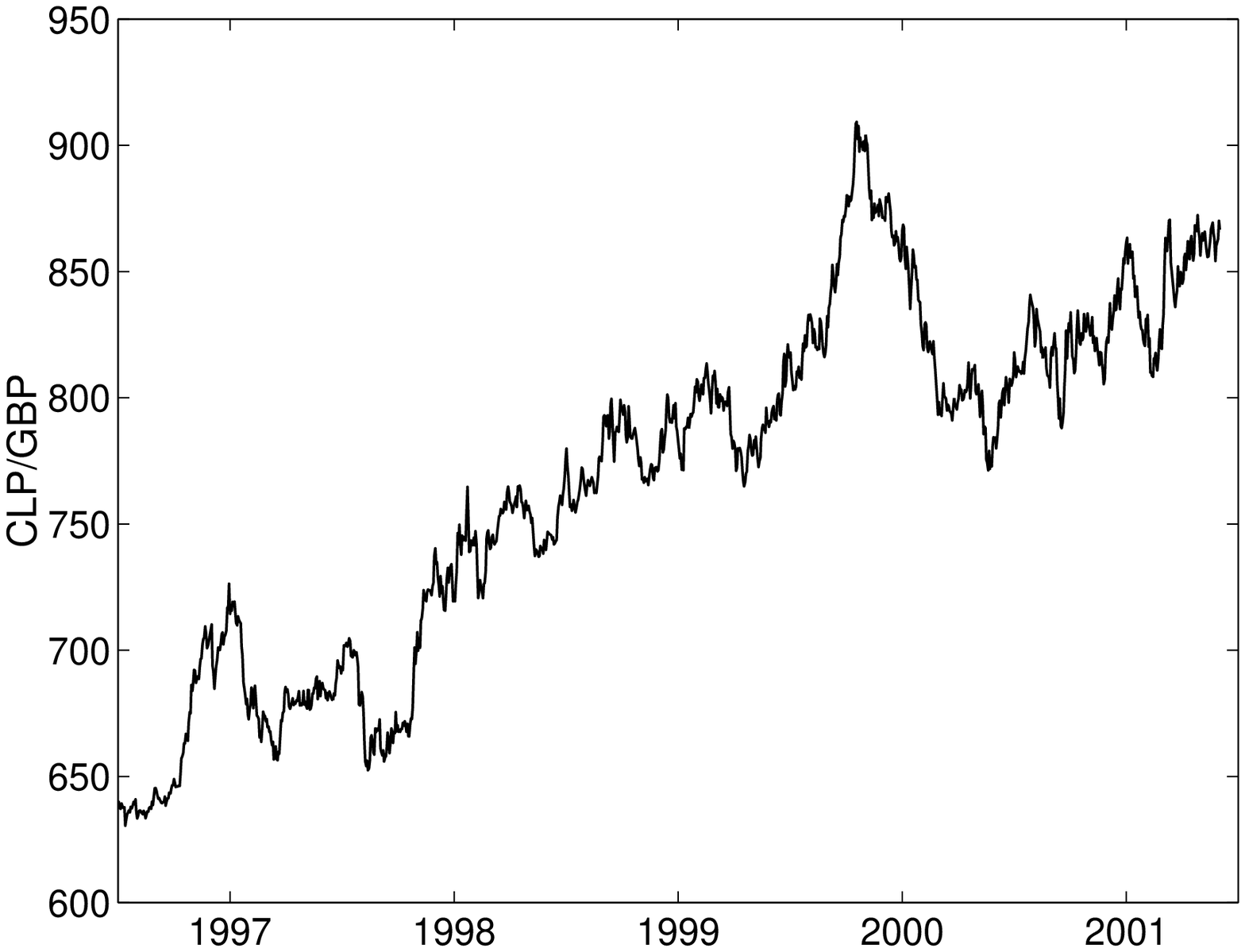} \vfill \caption[]{Exchange rates
of $CLP$ with respect to $SDR$, $USD$, $DEM$, $JPY$, $GBP$ for the time interval
between Nov. 16, 1995 and May 31, 2001, i.e. 1376 data points, as available on
http://pacific.commerce.ubc.ca/xr/ website } \label{eps7}\end{center} \end{figure}
\begin{figure} \begin{center}\leavevmode \epsfysize=3.3cm\epsffile{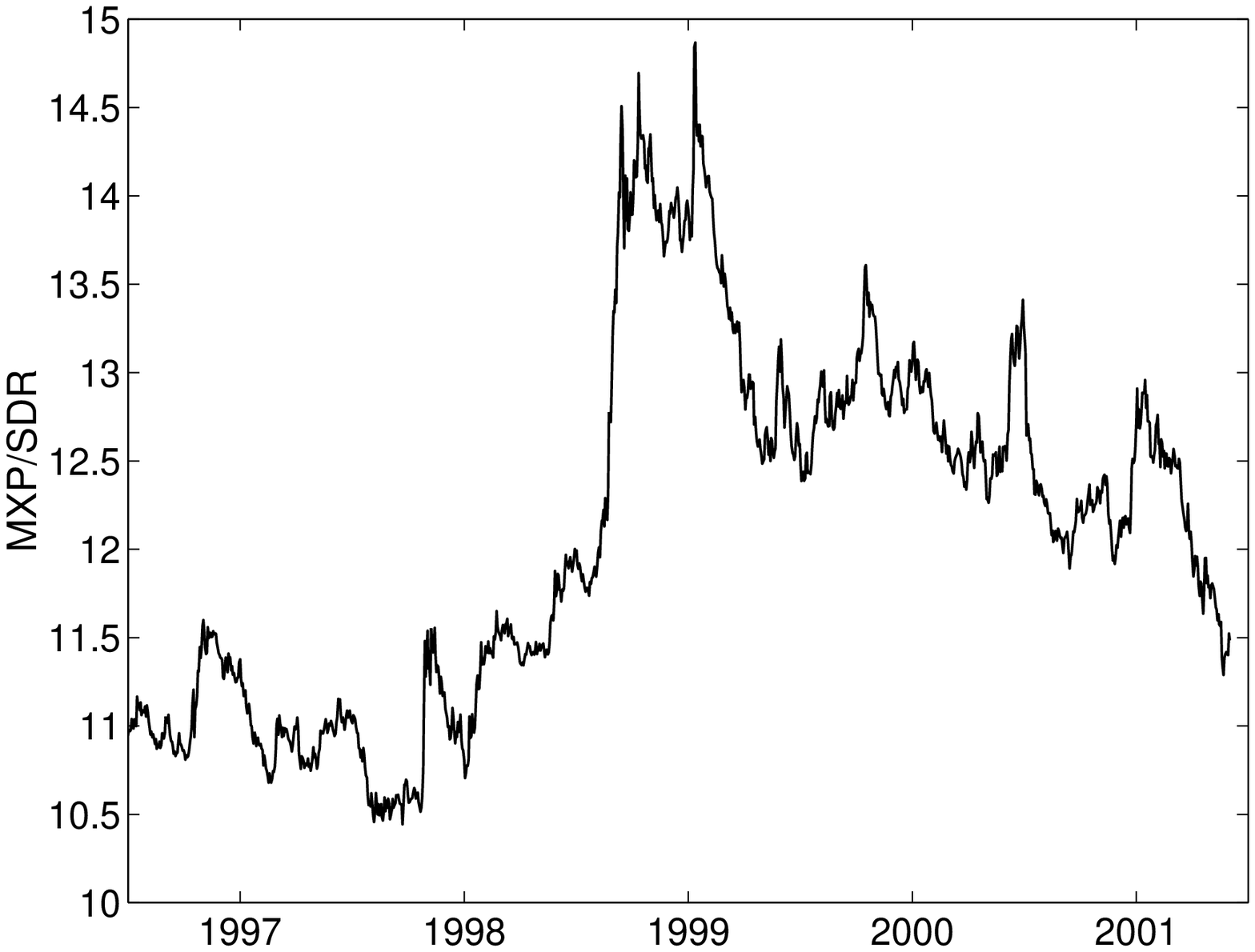} \hfill
\leavevmode \epsfysize=3.3cm\epsffile{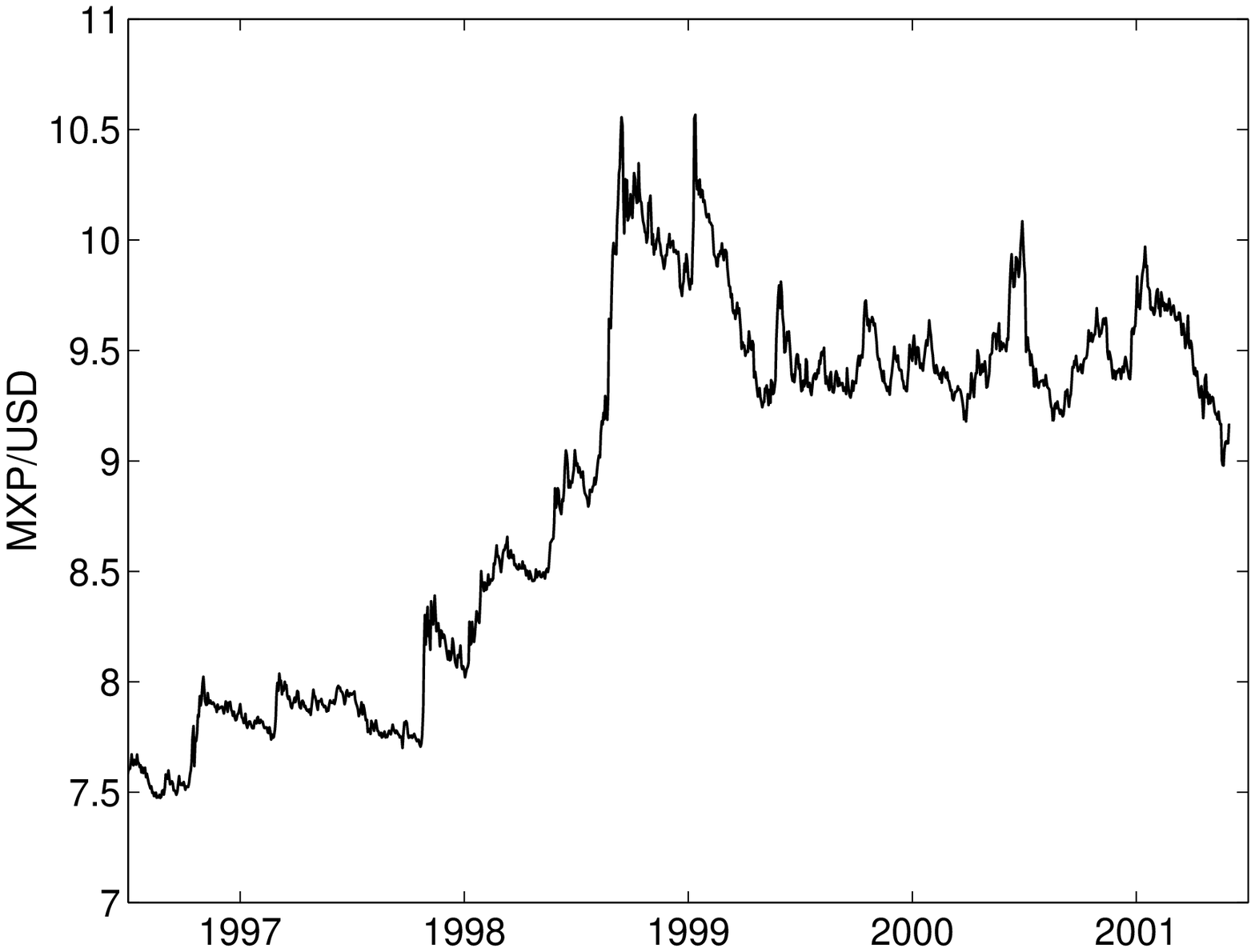} \hfill
\leavevmode \epsfysize=3.3cm\epsffile{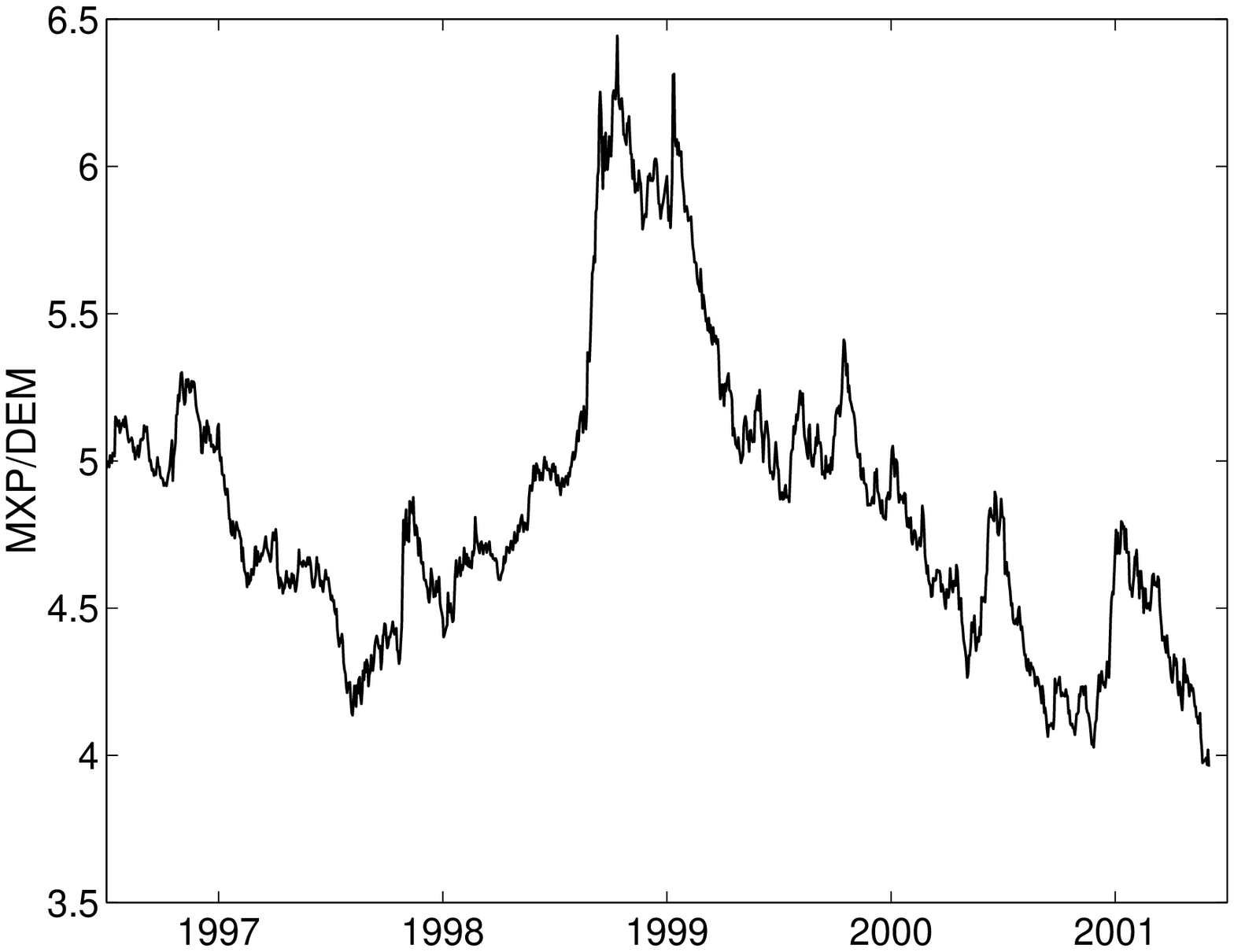} \vfill
\leavevmode \epsfysize=3.3cm\epsffile{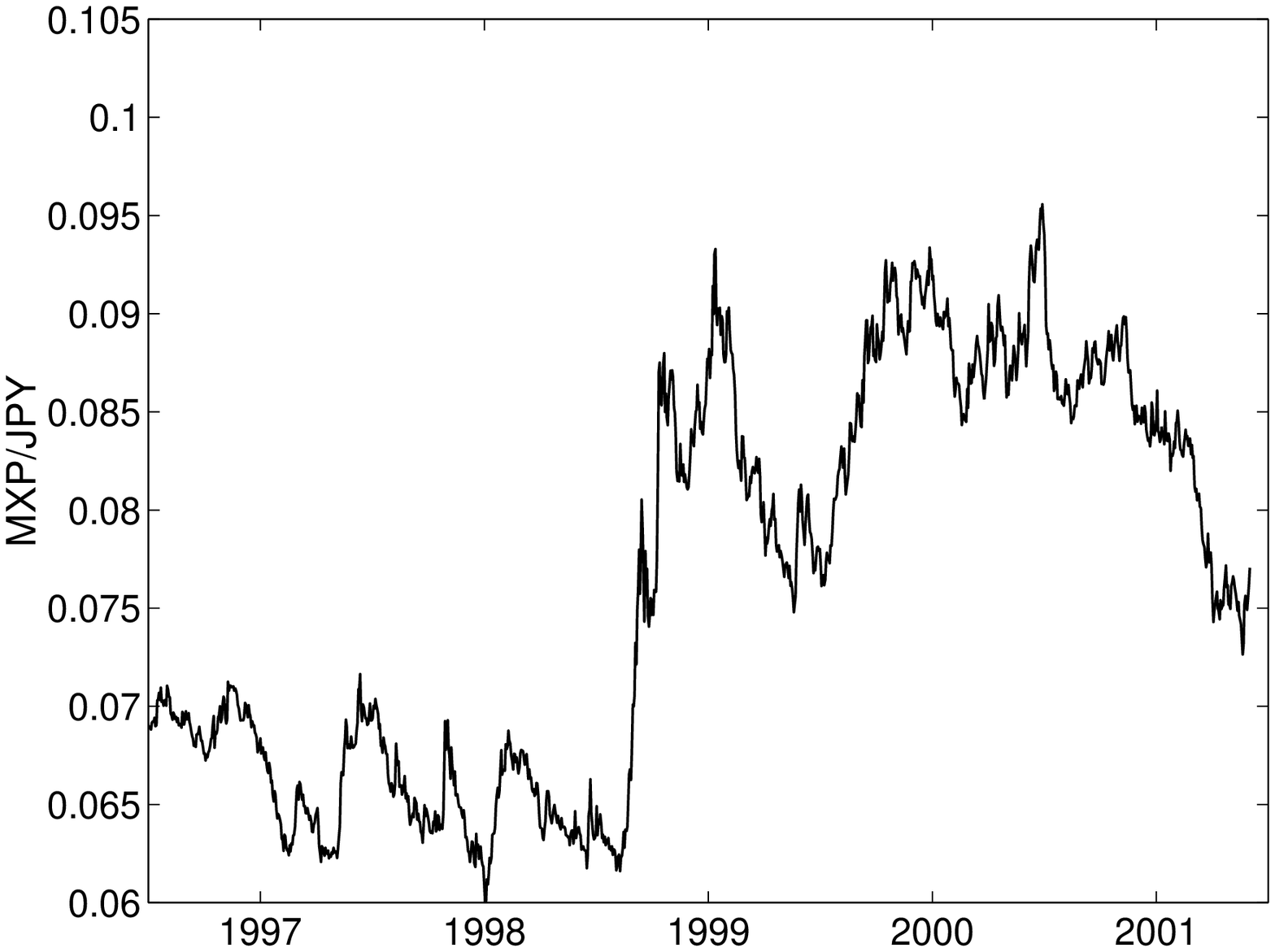} \hfill
\leavevmode \epsfysize=3.3cm\epsffile{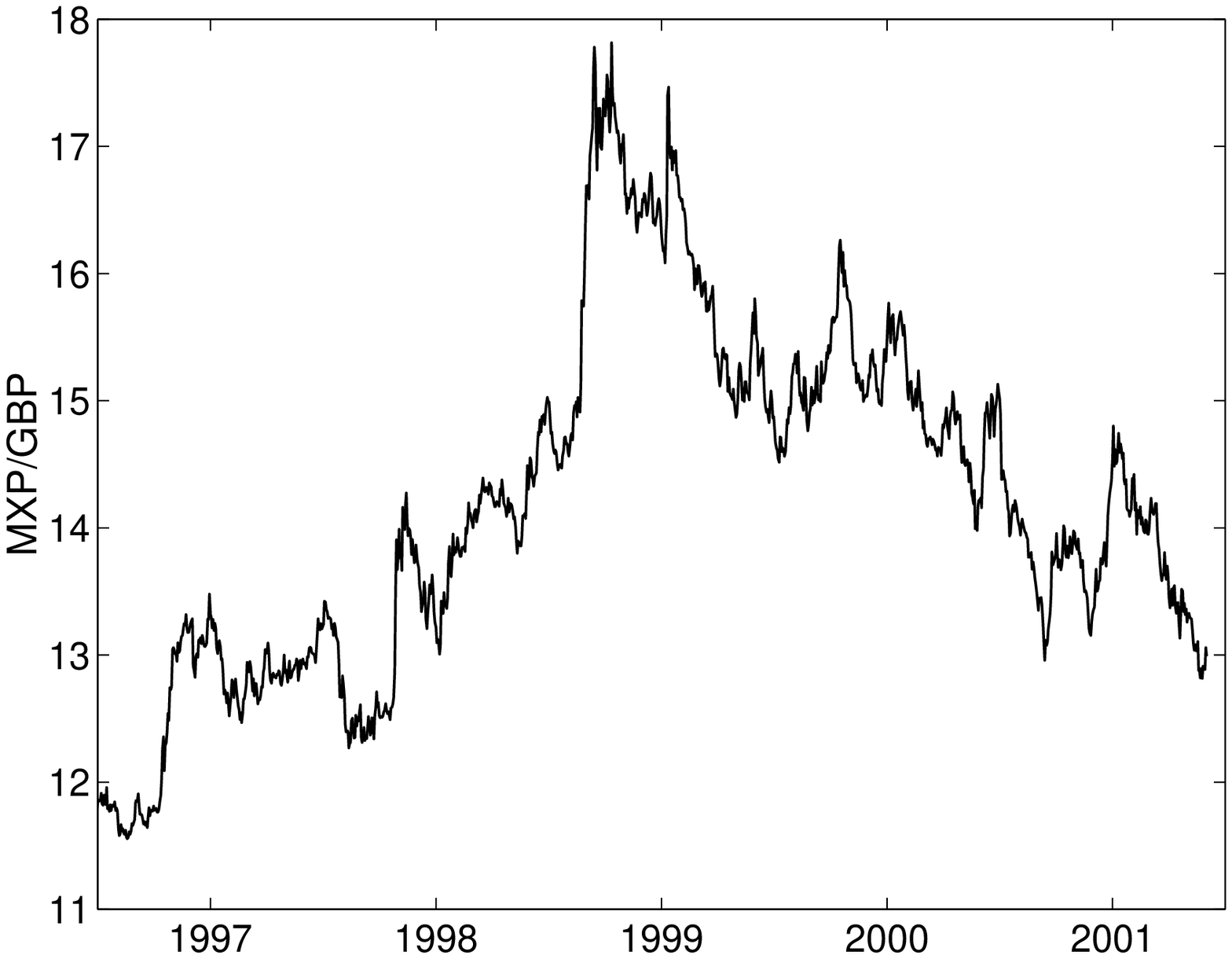} \vfill \caption[]{Exchange rates
of $MXP$ with respect to $SDR$, $USD$, $DEM$, $JPY$, $GBP$ for the time interval
between July 7, 1993 and May 31, 2001, i.e. 1975 data points, as available on
http://pacific.commerce.ubc.ca/xr/ website } \label{eps8} \end{center}\end{figure}

\subsection{The DFA technique}

The DFA technique \cite{DNADFA} is often used to study the correlations in the
fluctuations of stochastic time series like the currency exchange rates. Recall
briefly that the DFA technique consists in dividing a time series $y(t)$ of
length $N$ into $N/\tau$ nonoverlapping boxes (called also windows), each
containing $\tau$ points \cite{DNADFA}. The local trend $z(n)$ in each box is
defined to be the ordinate of a linear least-square fit of the data points in
that box. The detrended fluctuation function $F^2(\tau)$ is then calculated
following:

\begin{equation} F^2(\tau) = {1 \over \tau } {\sum_{n=k\tau+1}^{(k+1)\tau}
{\left[y(n)- z(n)\right]}^2} \qquad k=0,1,2,\dots,\left(\frac{N}{\tau}-1\right)
\end{equation}

Averaging $F^2(\tau)$ over the $N/\tau$ intervals gives the mean-square
fluctuations

\begin{equation} \phi(\tau)=<F^2(\tau)>^{1/2} \sim \tau^{\alpha} \end{equation}

The exponent $\alpha$ value implies the existence or not of long-range
correlations, and is assumed to be identical to the Hurst exponent when the data
is stationary. Moreover, $\alpha$ is an accurate measure of the most
characteristic (maximum) dimension of a multifractal process \cite{kiandma2}.
Since only the slopes and scaling ranges are of interest the various
DFA-functions $\phi(\tau)$ have been arbitrarily displaced for readability in
Figs. 9-11. The $\alpha$ values are summarized in Table 2. It can be noted that
the scaling ranges are usually from 5 days till 170 days for $ARS$ exchange
rates, from 5 days to about 1 year for $MXP$ and $CLP$ exchange rates, with the
exponent $\alpha$ close to 0.5 in that range. Crossover at ~80 days from Brownian
like to persistent correlations is obtained for $CLP/JPY$ and $CLP/SDR$.

\begin{figure}[ht] \begin{center}\leavevmode \epsfysize=9cm\epsffile{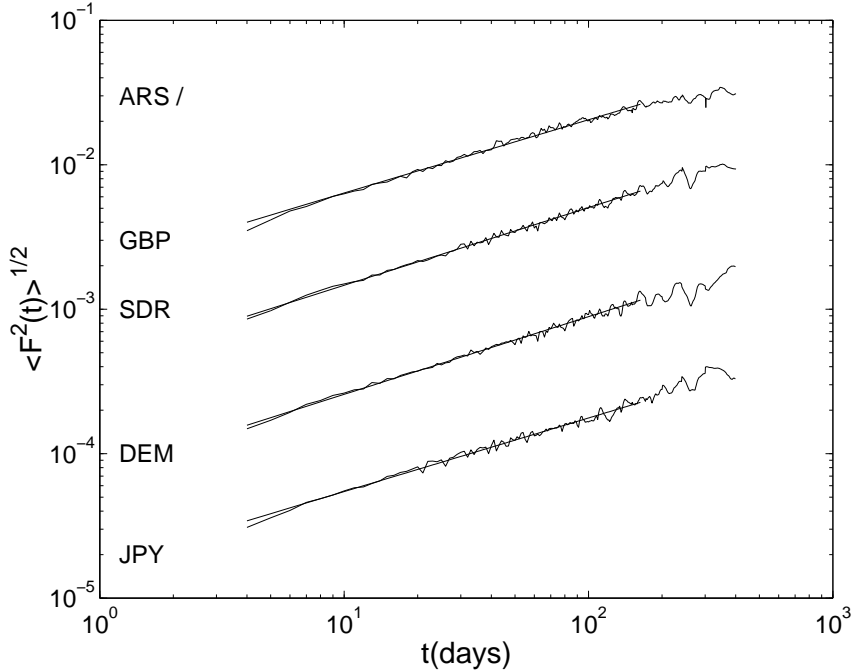}
\caption[]{DFA-function $\phi(\tau)$ function for the $ARS$ Exchange rates shown
in Fig. 6; the various $\phi(\tau)$ have been arbitrarily displaced for
readability} \label{eps9} \end{center}\end{figure}

\begin{figure}[ht] \begin{center}\leavevmode \epsfysize=9cm\epsffile{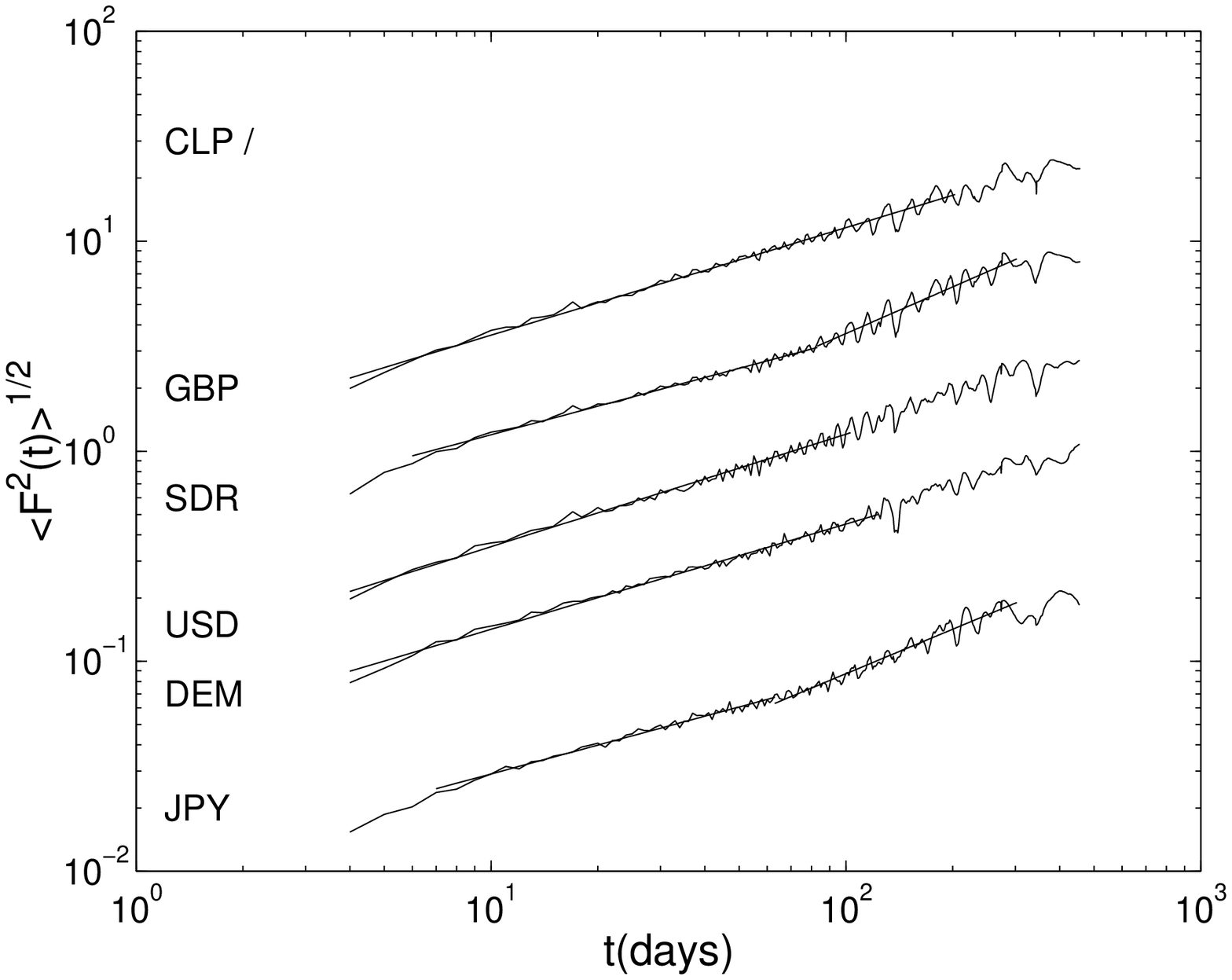}
\caption[]{DFA-function $\phi(\tau)$ function for the $CLP$ Exchange rates shown
in Fig. 7; the various $\phi(\tau)$ have been arbitrarily displaced for
readability} \label{eps10} \end{center}\end{figure}

\begin{figure}[ht] \begin{center}\leavevmode \epsfysize=9cm\epsffile{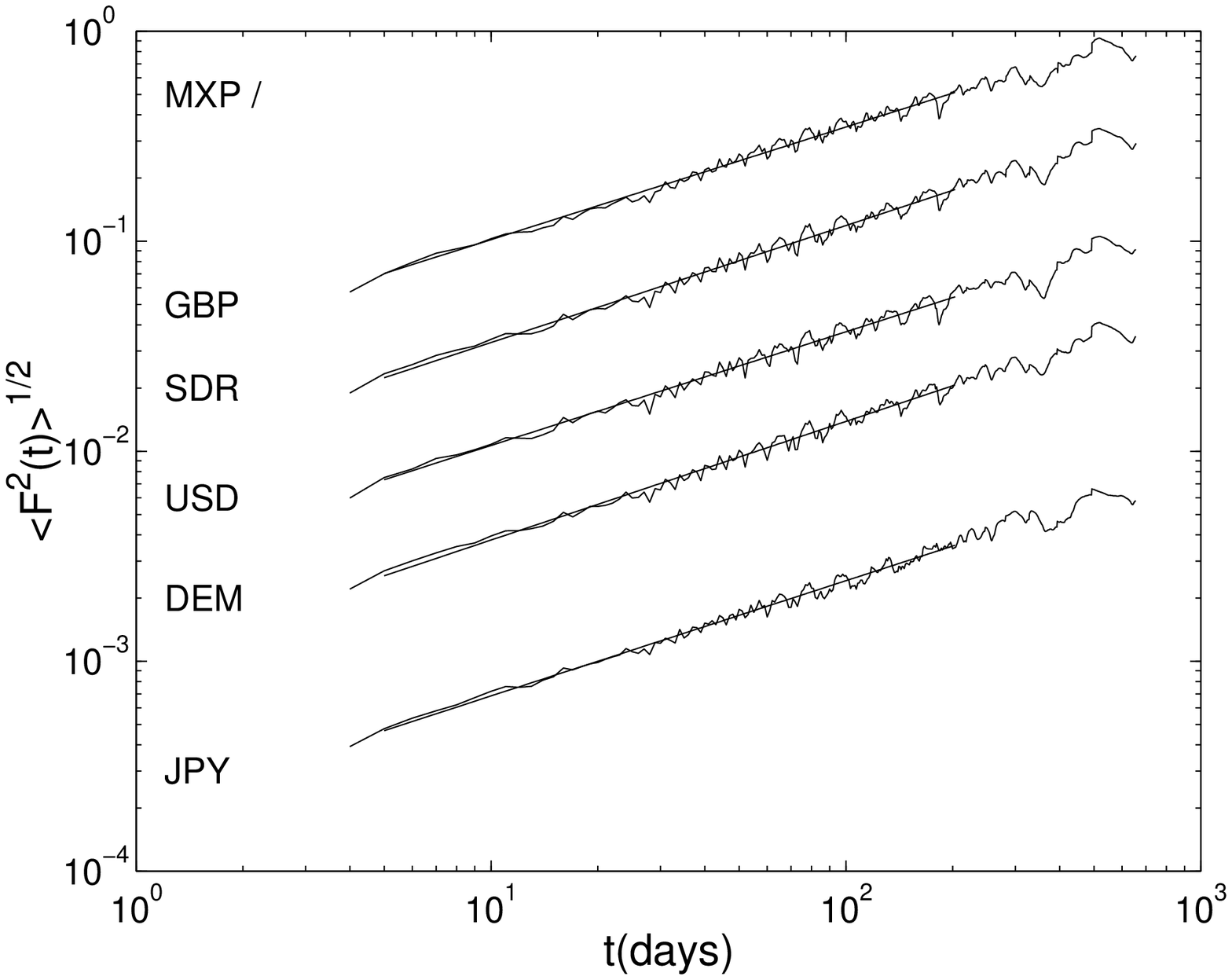}
\caption[]{DFA-function $\phi(\tau)$ function for the $MXP$ Exchange rates shown
in Fig. 8; the various $\phi(\tau)$ have been arbitrarily displaced for
readability} \label{eps11} \end{center}\end{figure}

\begin{table}[htb] \begin{center} \caption{$\alpha$ exponent for the scaling
regime of considered ExR in the text} \begin{tabular}{|c|c|c|c|c|c|} \hline
&$USD$&$DEM$&$GBP$&$JPY$&$SDR$\\[6pt] \hline $ARS /$&&
0.54$\pm$0.02&0.51$\pm$0.02&0.51$\pm$0.02&0.54$\pm$0.02\\[6pt] \hline $CLP
/$&0.54$\pm$0.03&0.50$\pm$0.02&0.51$\pm$0.03&0.46$\pm$0.02&0.45$\pm$0. 01\\[6 pt]
&&&&0.70$\pm$0.07&0.74$\pm$0.07\\[6pt] \hline $MXP
/$&0.54$\pm$0.03&0.56$\pm$0.03&0.54$\pm$0.03&0.55$\pm$0.02&0.56$\pm$0. 03\\
\hline \end{tabular} \end{center} \label{table2}\end{table}

\subsection{Local scaling with DFA and Intercorrelations between fluctuations}

The time derivative of $\alpha$ can usually be correlated to an {\it entropy}
production \cite{nvma} through market information exchanges. As done elsewhere, in
order to probe the existence of {\it locally correlated} and {\it decorrelated}
sequences, we have constructed an observation box, i.e. a 500 days wide window
probe placed at the beginning of the data, calculated $\alpha$ for the data in
that box. Moving this box by one day toward the right along the signal sequence
and again calculating $\alpha$, a local $\alpha$ exponent is found (but not
displayed here).

\begin{figure}[ht] \begin{center}\leavevmode \epsfysize=3.5cm\epsffile{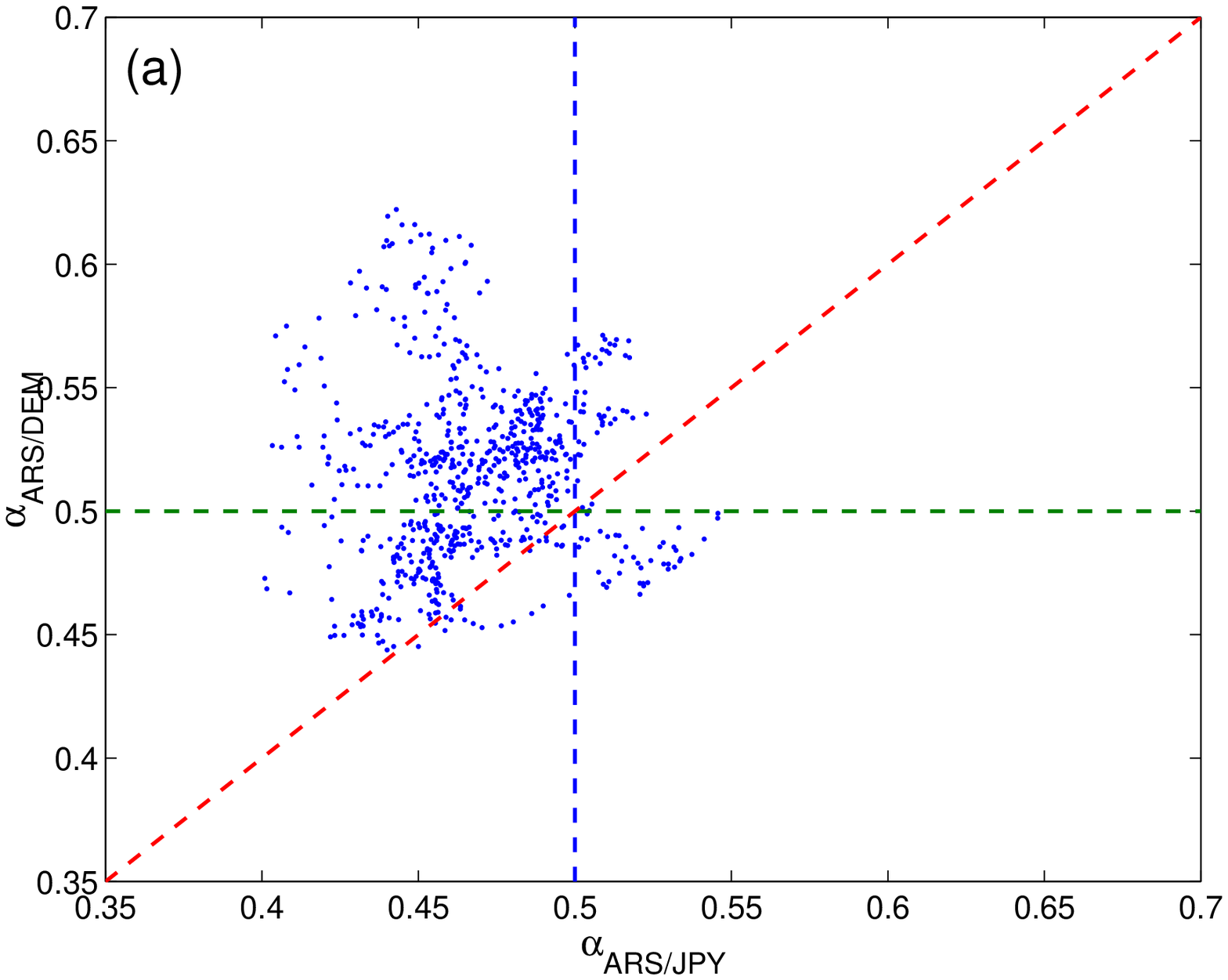}
\hfill \leavevmode \epsfysize=3.5cm\epsffile{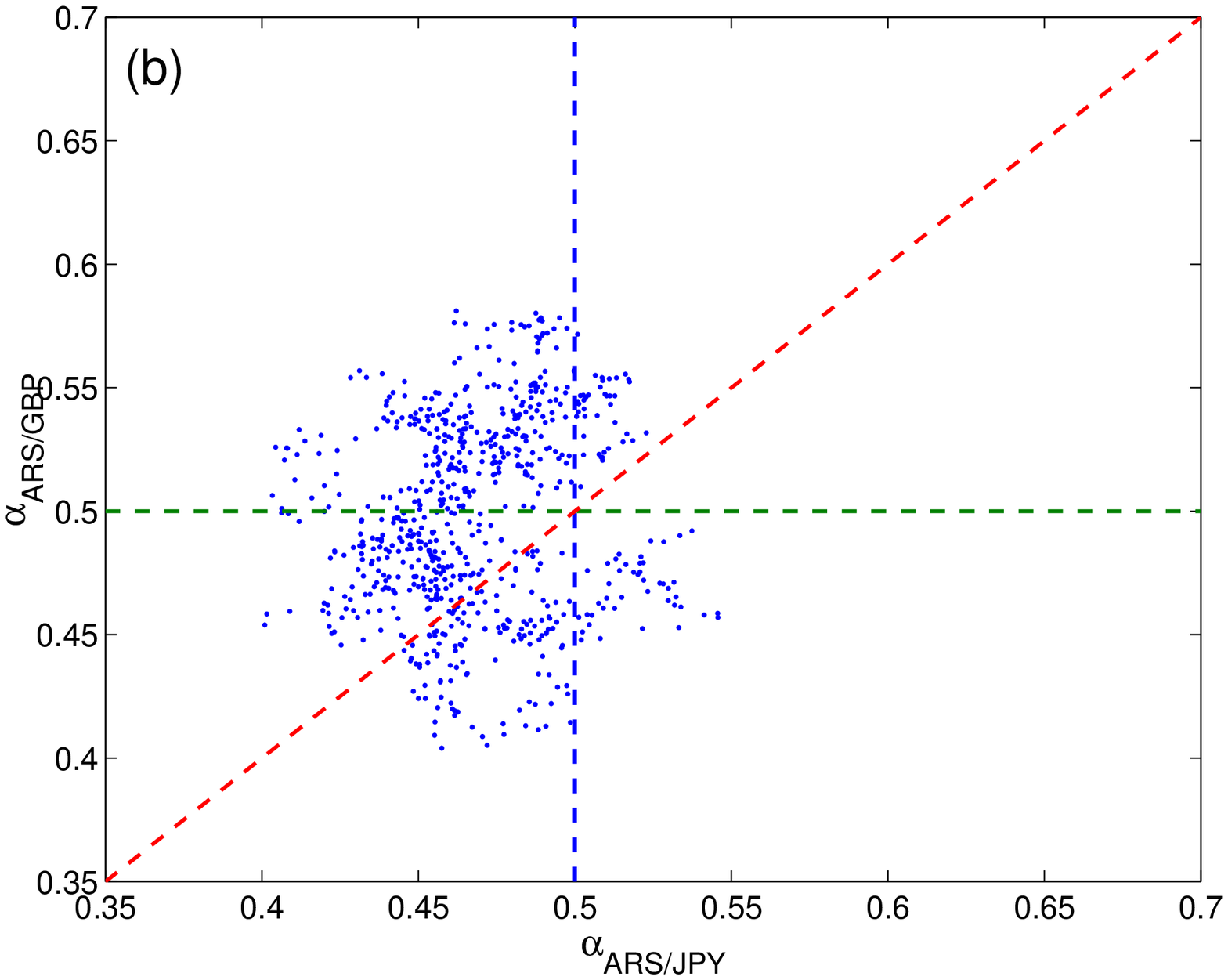} \hfill
\leavevmode \epsfysize=3.5cm\epsffile{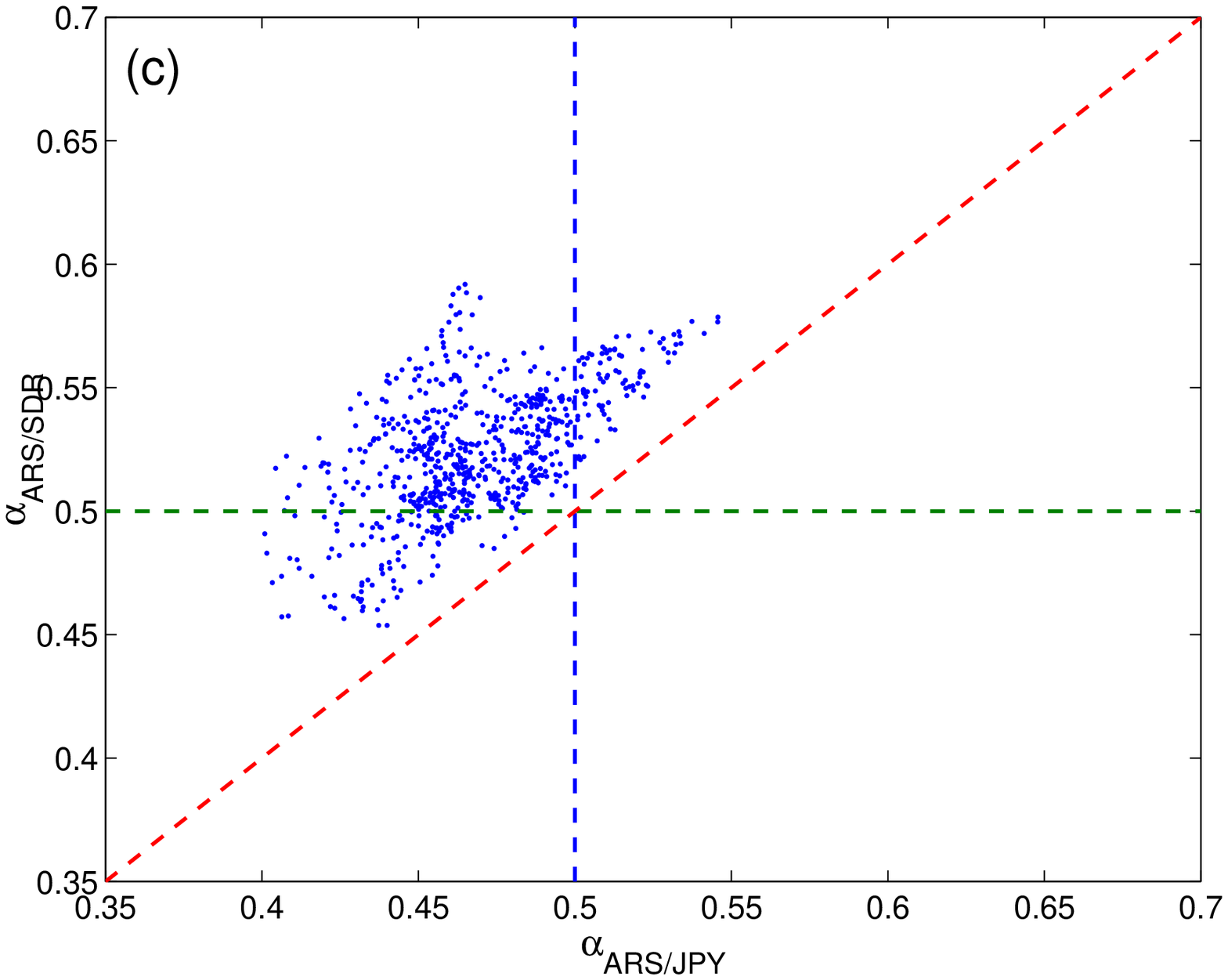} \vfill
\leavevmode \epsfysize=3.5cm\epsffile{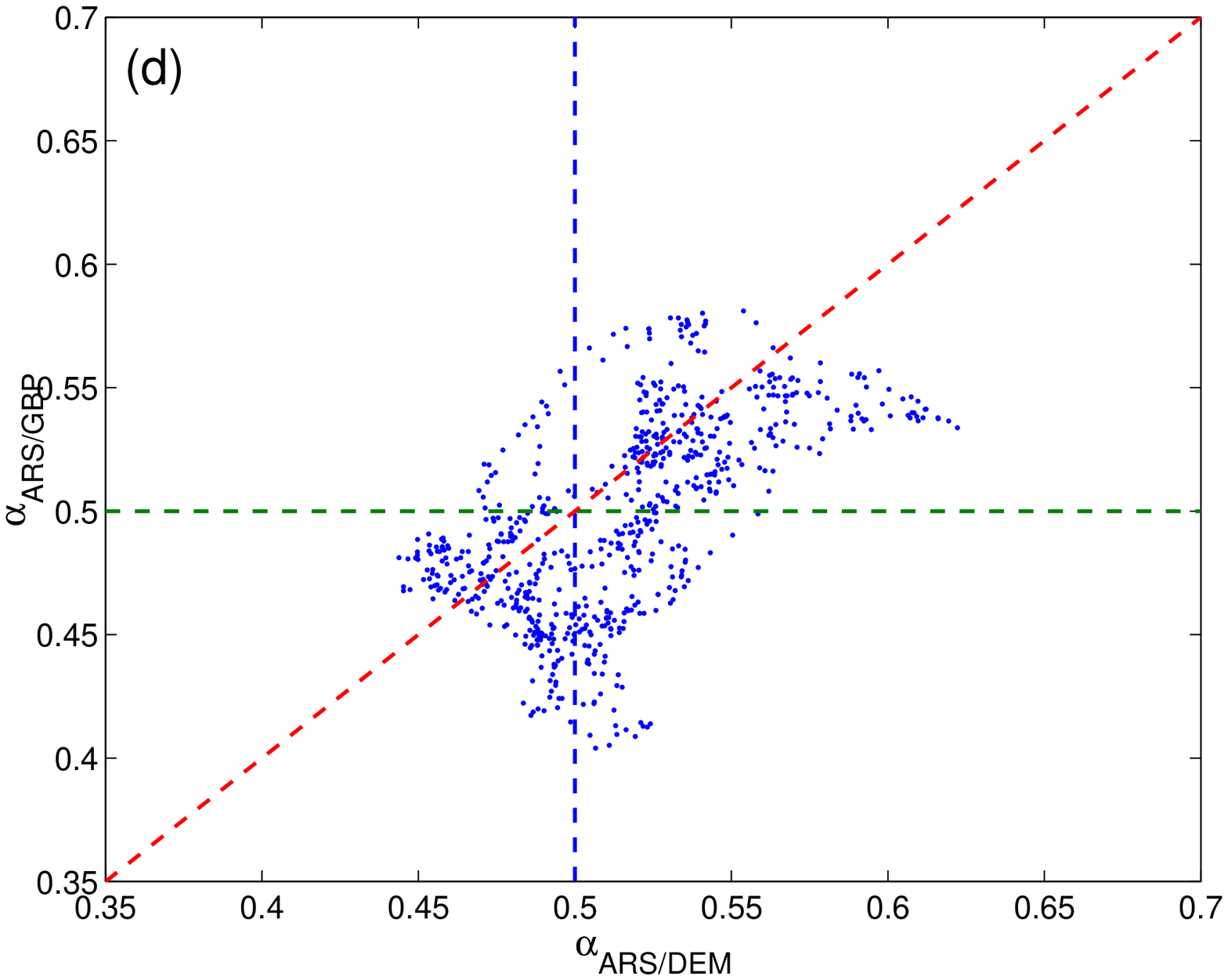} \hfill
\leavevmode \epsfysize=3.5cm\epsffile{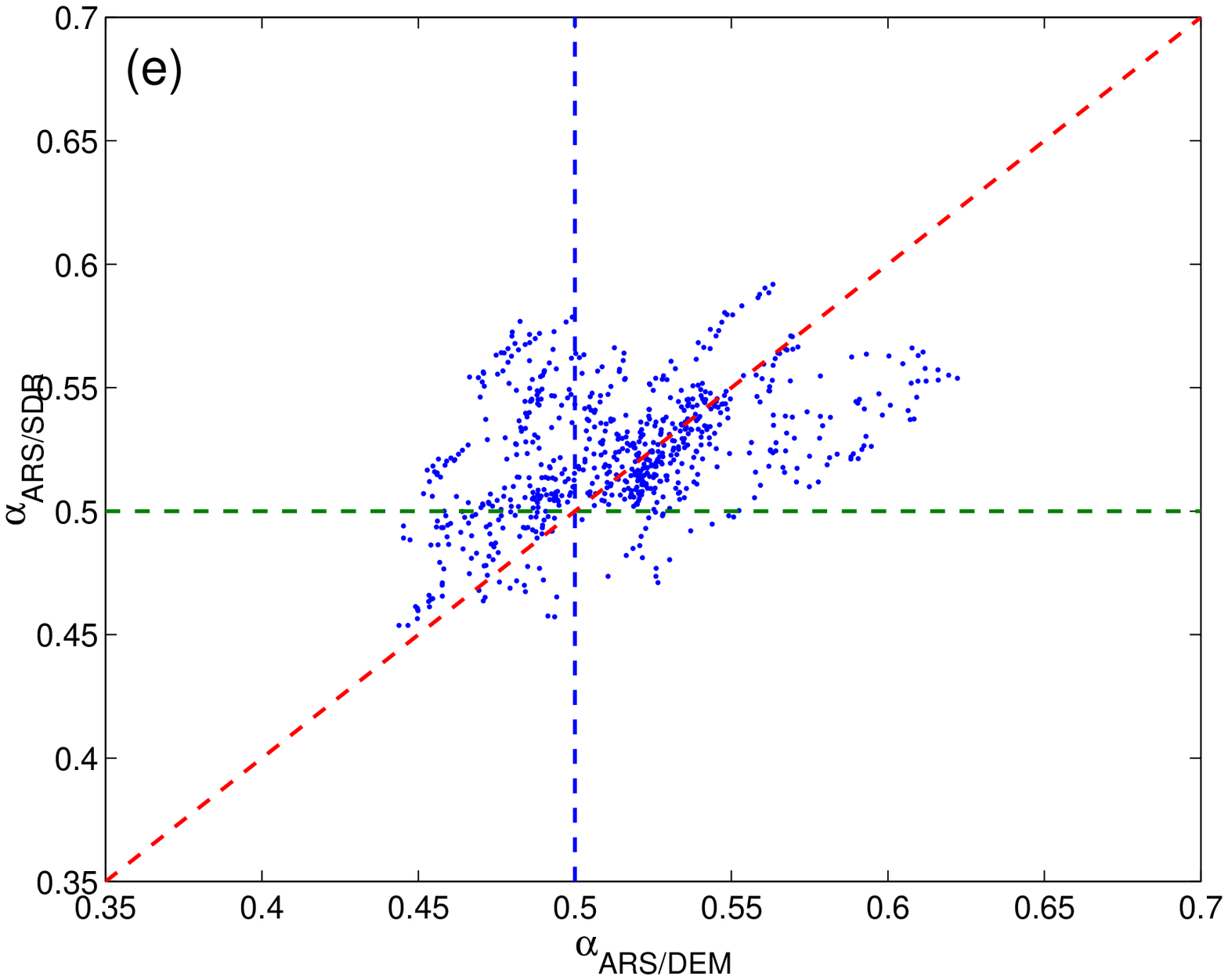} \hfill
\leavevmode \epsfysize=3.5cm\epsffile{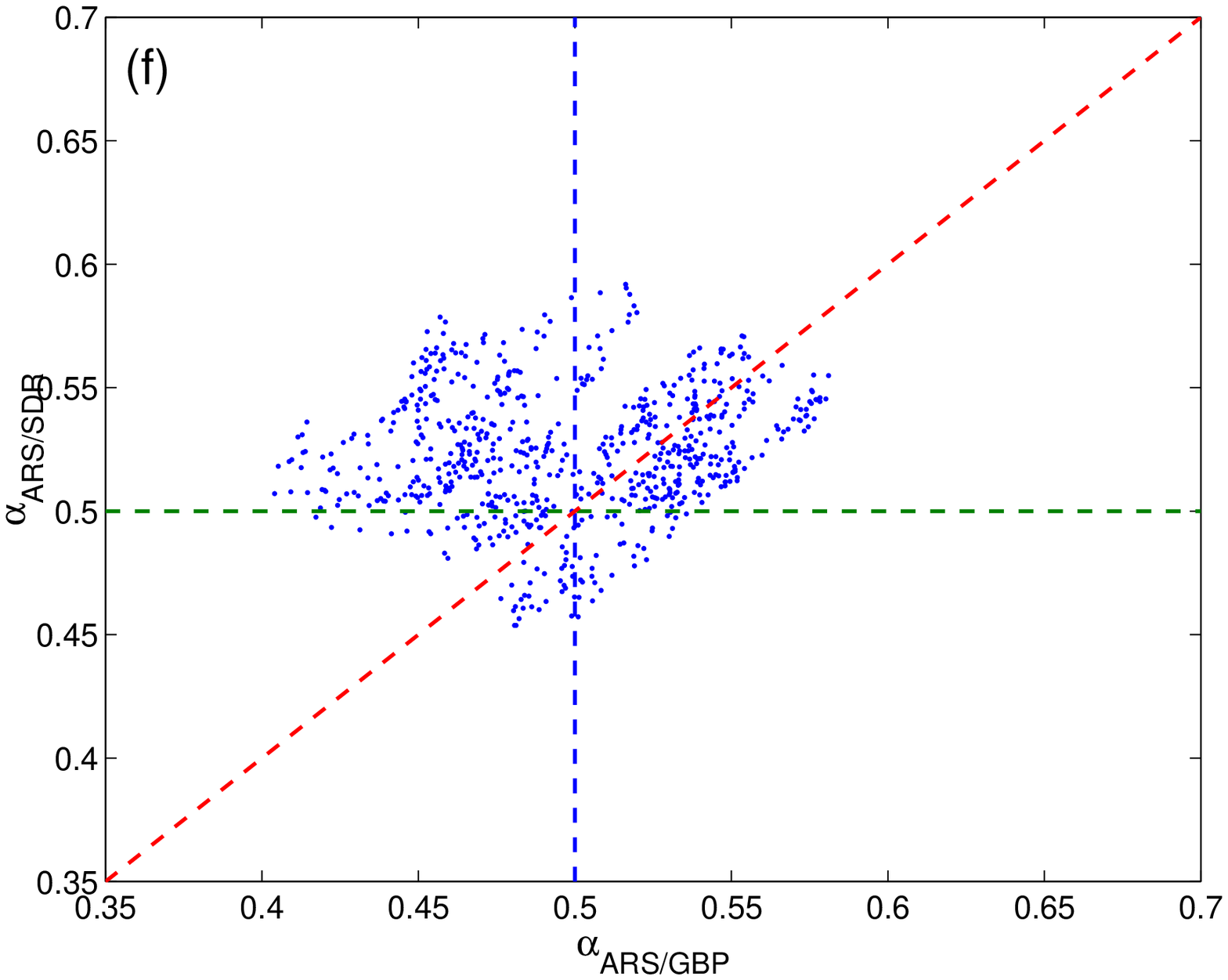}  \caption[]{Structural
correlation diagram of (a-f) between typical $\alpha_{C_i/B_j}$ exponents for
exchange rates between $ARS$ and $DEM$, $GBP$, $JPY$, $SDR$ }\label{eps12}
\end{center}\end{figure}

\begin{figure}[ht] \begin{center}\leavevmode \epsfysize=3.5cm\epsffile{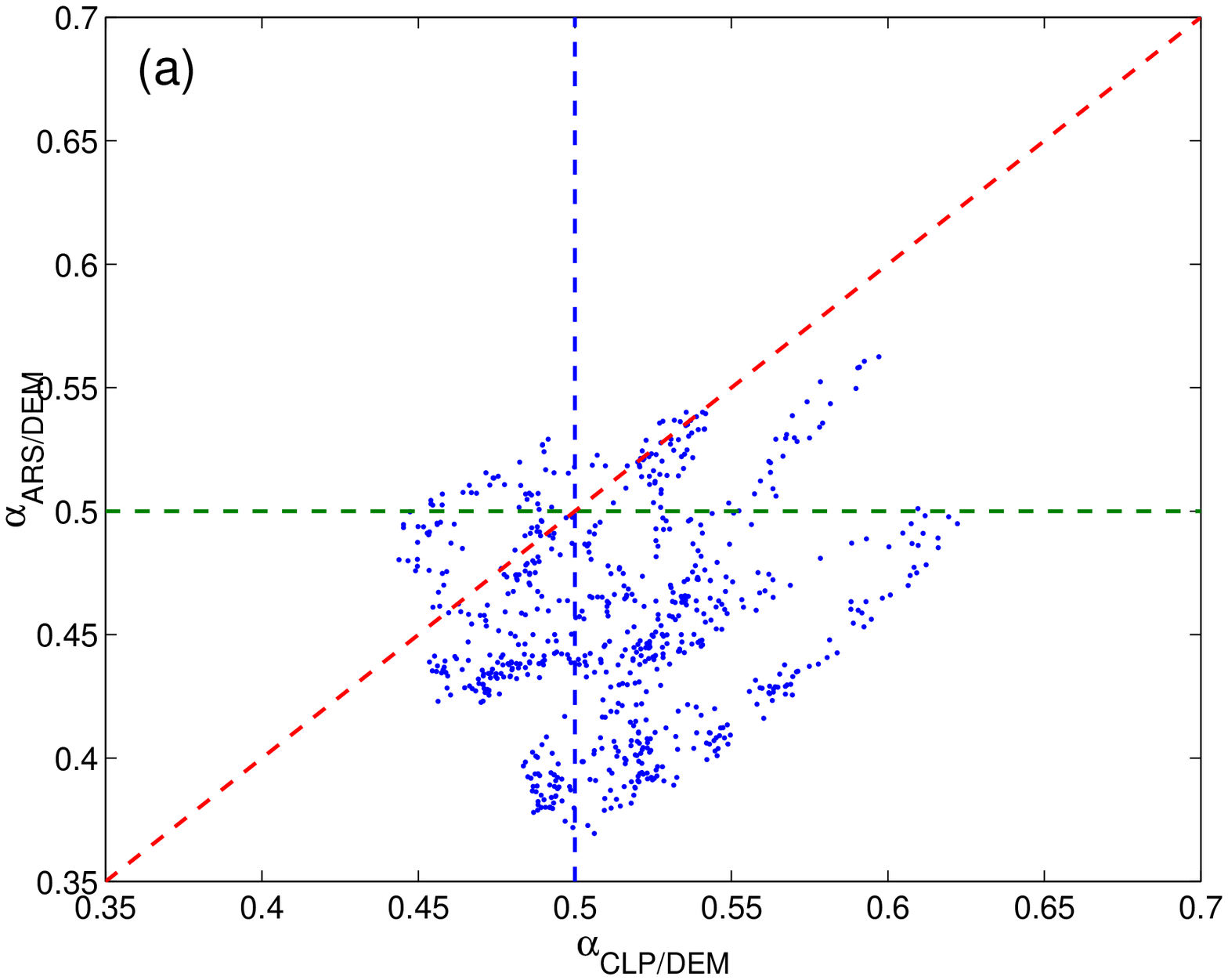}
\hfill \leavevmode \epsfysize=3.5cm\epsffile{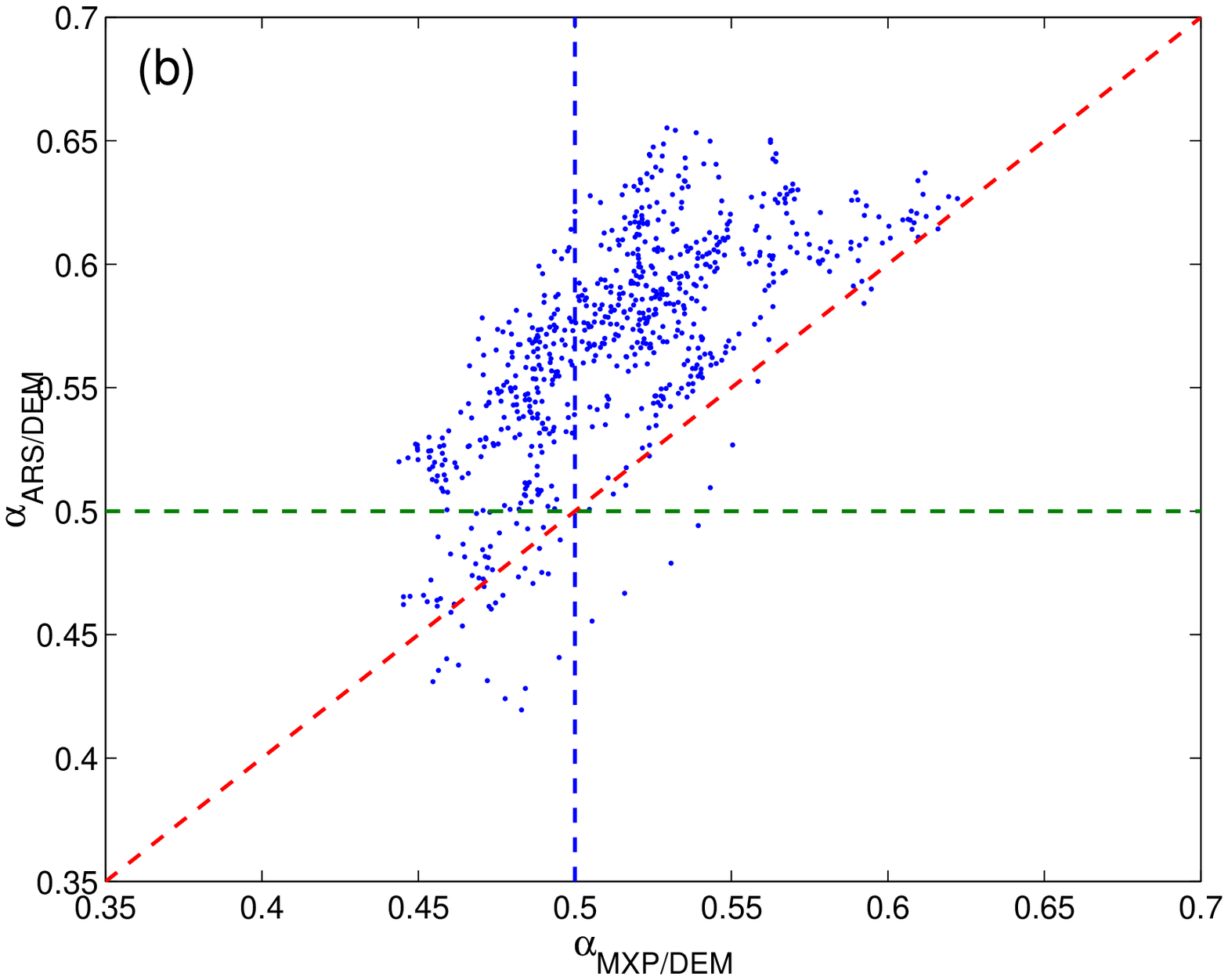} \hfill
\leavevmode \epsfysize=3.5cm\epsffile{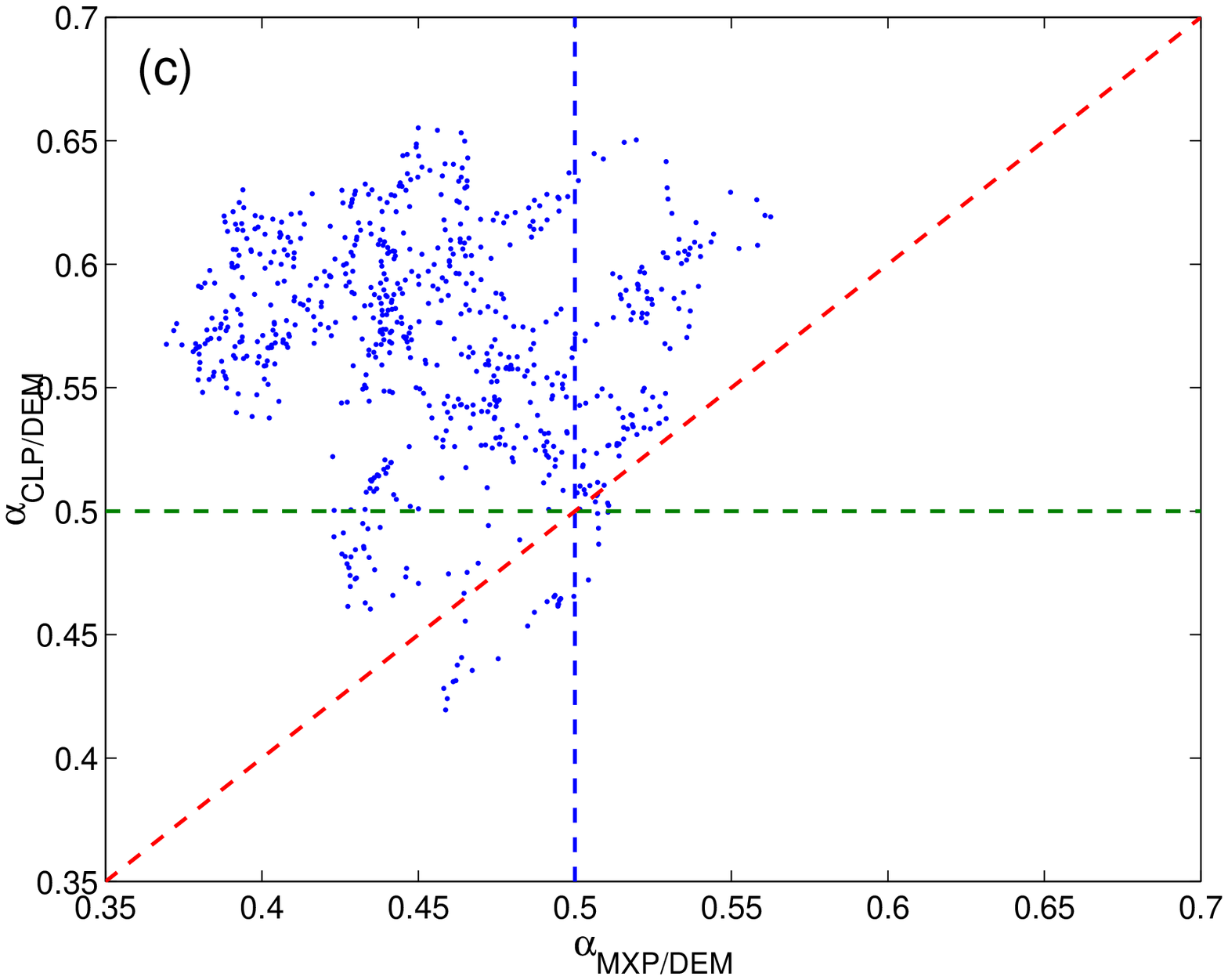} \caption[]{Structural
correlation diagram of (a-c) between typical $\alpha_{C_i/B_j}$ exponents for
exchange rates, i.e. involving $ARS$, $MXP$, $CLP$ and $DEM$ }\label{eps13}
\end{center}\end{figure}

\begin{figure}[ht] \begin{center}\leavevmode \epsfysize=3.5cm\epsffile{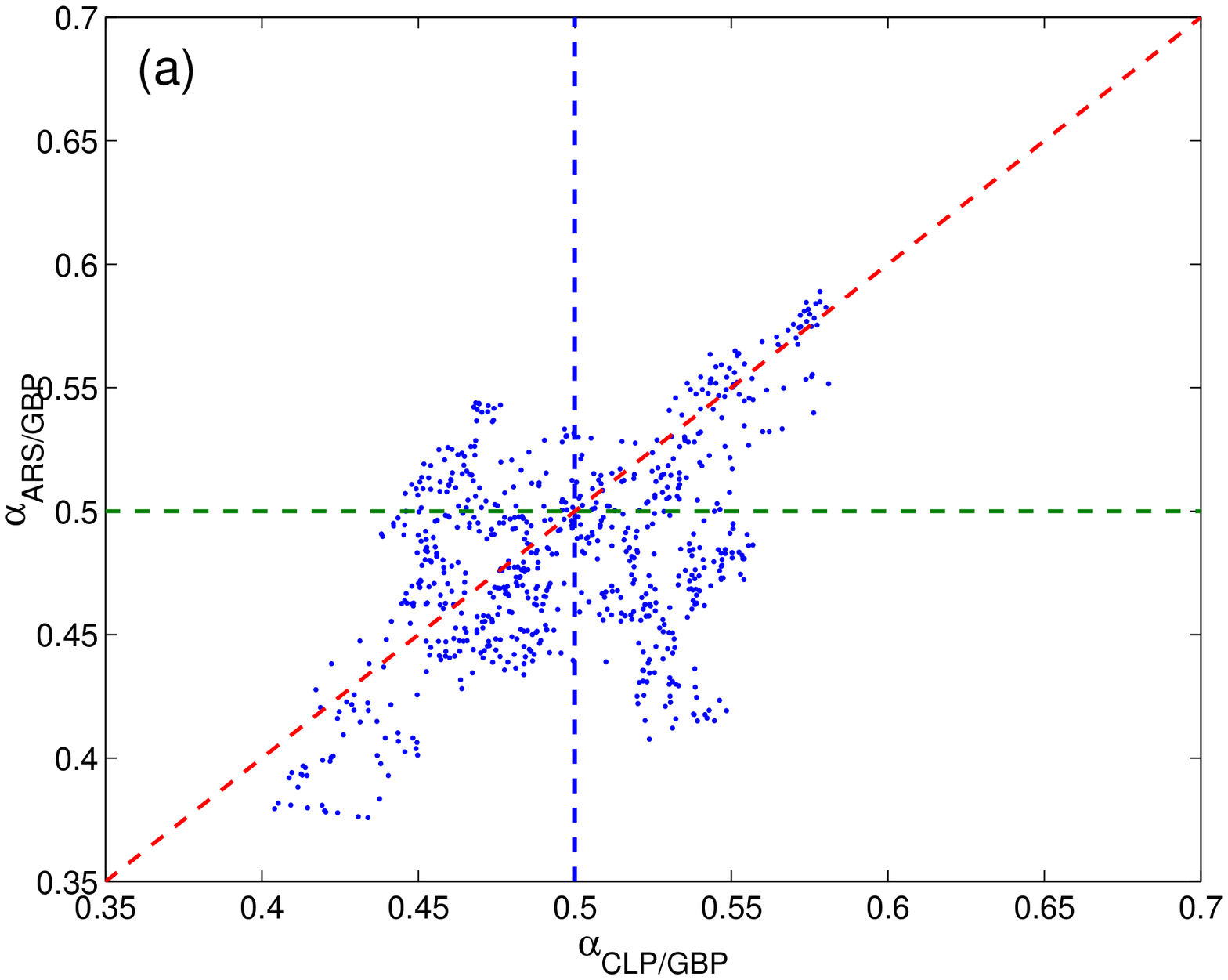}
\hfill \leavevmode \epsfysize=3.5cm\epsffile{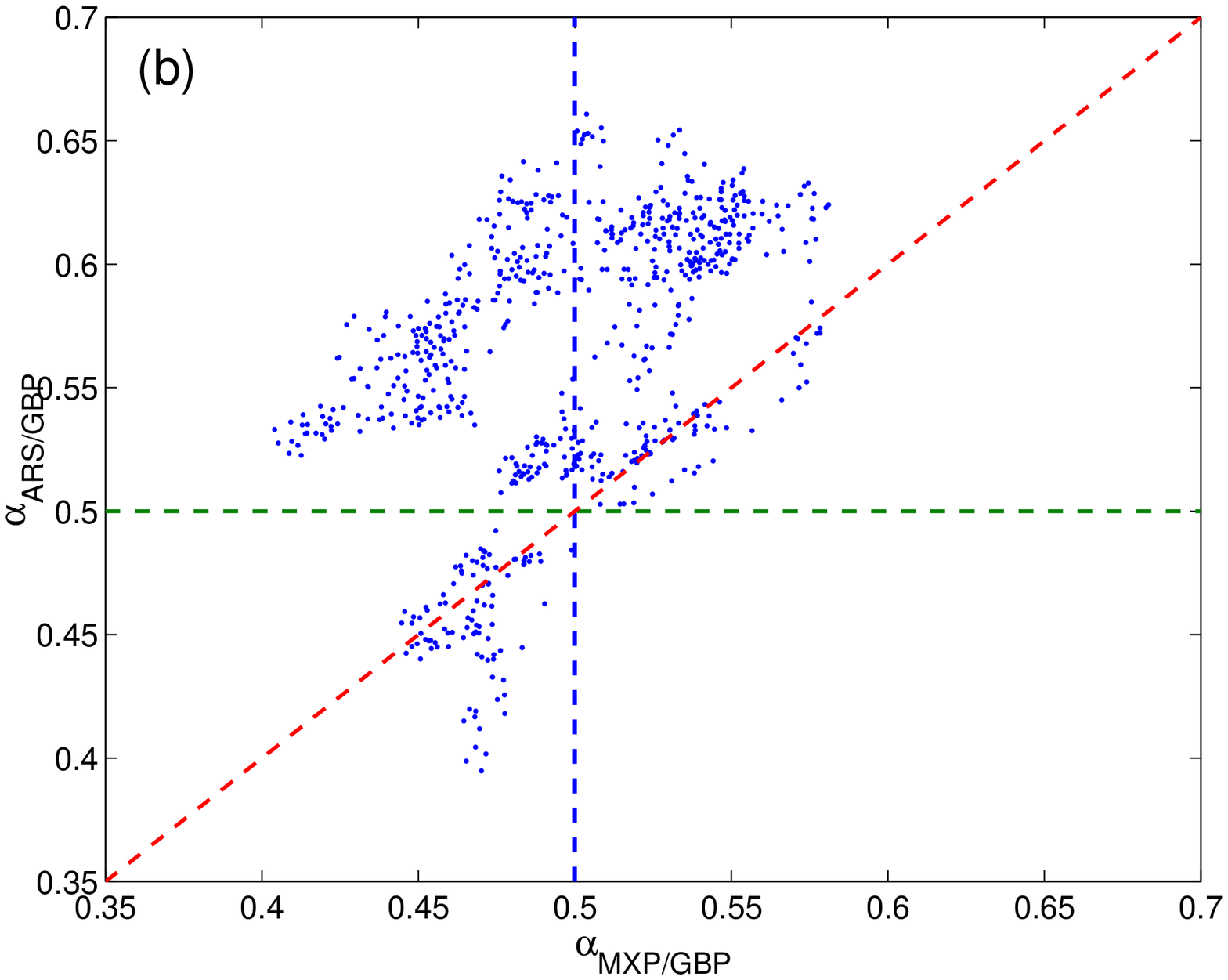} \hfill
\leavevmode \epsfysize=3.5cm\epsffile{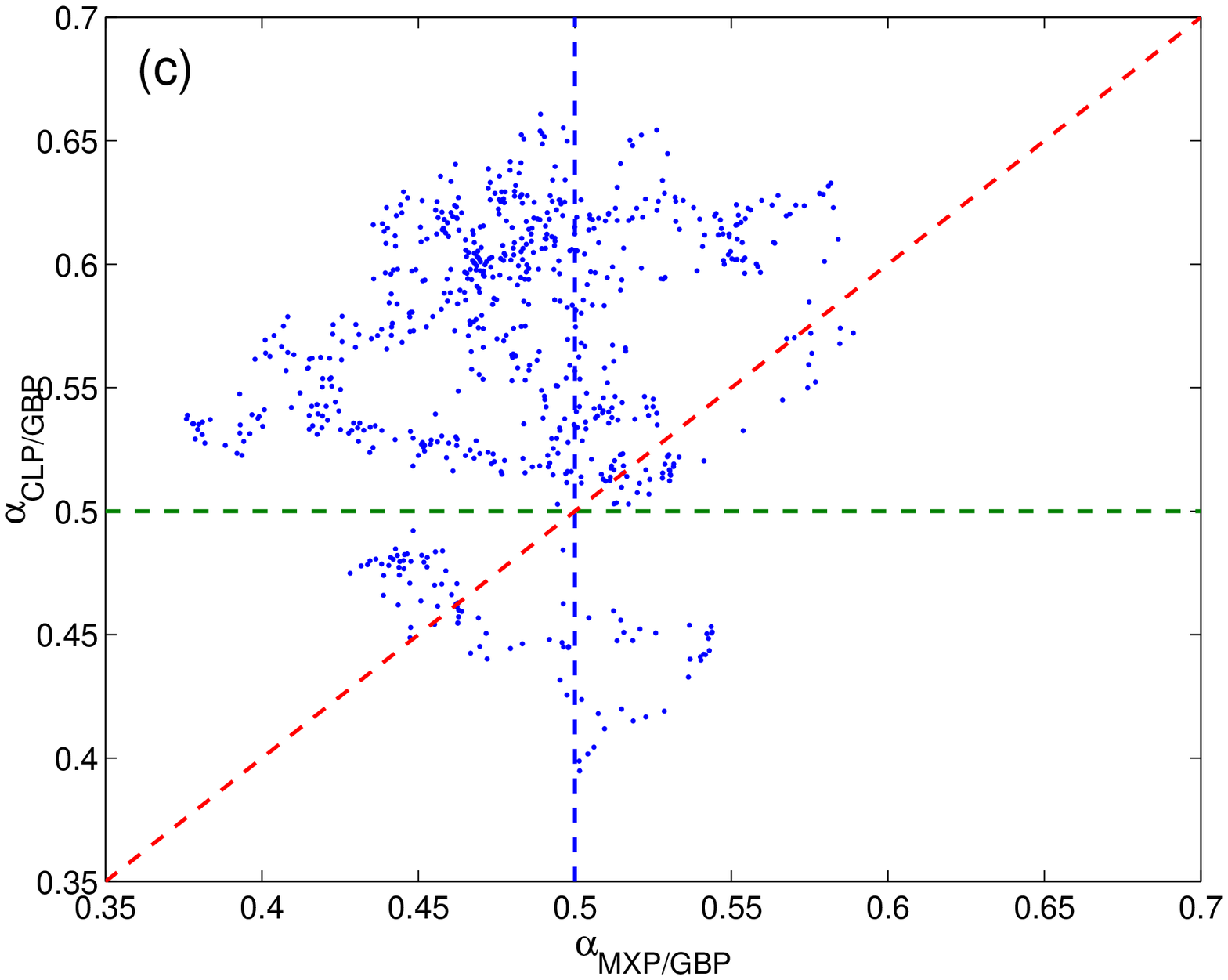} \caption[]{Structural
correlation diagram of (a-c) between typical $\alpha_{C_i/B_j}$ exponents for
exchange rates, i.e. involving $ARS$, $MXP$, $CLP$ and $GBP$ }\label{eps14}
\end{center}\end{figure}

We eliminate the time between these data sets and construct a graphical
correlation matrix of the time-dependent $\alpha$ exponents for the various
exchange rates of interest (Fig.12-14). We show $\alpha_{C_i/B_j}$ {\it vs.}
another $\alpha_{C_i/B_j}$, where a $C_i$ is a developing country currency while
$B_j$ is $USD$, $GBP$, $JPY$, $DEM$, and $SDR$ for the available data. In so
doing a so-called correlation matrix is displayed for the time interval of
interest. Such bilateral correlations between different $\alpha$ exponents can be
considered in order to estimate the strength and some nature of the correlations.
As described elsewhere \cite{kimalg}, such a correlation diagram can be divided
into main sectors through a horizontal, a vertical and diagonal lines crossing at
(0.5,0.5). If the correlation is strong the cloud of points should fall along the
slope $= + 1$ line. If there is no correlation the cloud should be rather
symmetrical. The lack of symmetry of the plots and wide spreading of points
outside expected clouds (see e.g. Figs. 13(c), 14(b,c)  - mainly containing
$MXP$) indicate highly speculative situations.  Notice the marked imbalance of
some plots, mainly involving $ARS$. It is fair to say that  other techniques are
also of great interest to observe correlations between  financial markets
\cite{canal,mansilla,maslov}.

\section{Conclusions}

The classical technical analysis methods of financial indices, stocks, futures,
... are very puzzling. We have recalled them. Illustrations have used the IBM share
price and Latin American financial indices. We have used the DFA method
to search for scaling ranges and type of behavior of exchange rates
between Latin American currencies ($ARS$, $CLP$, $MXP$) and other major
currencies $DEM$, $GBP$, $JPY$ and $USD$, including $SDR$s.
In all cases persistent to Brownian like behavior is obtained for 
scaling ranges from a week to about one year, with an exception of 
$CLP/JPY$ and $CLP/SDR$ for which there is a transition from Brownian
like to persistent correlations with $\alpha=0.70$ and $\alpha=0.74$ 
for scaling ranges longer than 80~days. We have also sorted
out to correlations and anticorrelations of such exchange rates with respect to
currencies as $DEM$, $GBP$, $JPY$ and $USD$. They indicate a very complex or
speculative behavior. 

\vskip 2cm

\noindent {\bf Acknowledgements}

MA thanks to the organizers of the Stauffer 60th birthday symposium for their
invitation and kind welcome.

 \end{document}